\documentstyle[12pt,newpp4]{article}

\lefthead{}
\righthead{}

\begin{document}

\null

{\vskip -2.2truecm\hskip 5.5truecm
  \vtop{\hsize=10.truecm \hyphenpenalty=5000 \noindent
    To appear in {\it Annual Reviews of Astronomy \break \& Astrophysics}, 1999 }}

\vskip2.0truecm


\def\mathfont#1{\ifmmode{#1}\else{$#1$}\fi}

\def\la{\mathrel{\hbox{\rlap{\hbox{\lower4pt\hbox{$\sim$}}}\hbox{$<$}}}}
\def\ga{\mathrel{\hbox{\rlap{\hbox{\lower4pt\hbox{$\sim$}}}\hbox{$>$}}}}

\def\flun{\mathfont{ {\rm erg\ s}^{-1}\ {\rm cm}^{-2}\ {\rm
\AA}^{-1}}}

\def\msun{\mathfont{{\rm M}_\odot}}
\def\msunyr{\mathfont{\msun \ {\rm yr}^{-1}}}
\def\rsun{\mathfont{ {\rm R}_\odot}}

\def\lmet{\mathfont{\log Z/Z_\odot}}
\def\feh{\mathfont{[Fe/H]}}
\def\htwo{H$_2$}
\def\teff{\mathfont{T_e}}
\def\lbol{\mathfont{L_{bol}}}
\def\mv{\mathfont{m_V}}
\def\MV{\mathfont{M_V}}
\def\ergsec{\mathfont{ {\rm ergs\ s}^{-1}}}
\def\ergcmsec{\mathfont{ {\rm erg\ s}^{-1}\ {\rm cm}^{-2}}}

\def\efifteen{\mathfont{E_{1500}}}
\def\zsun{\mathfont{Z_{\odot}}}
\def\etar{\mathfont{\eta_R}}
\def\menv{\mathfont{M_{ENV}}}

\def\lam{\mathfont{\lambda}}
\def\lamlam{$\lambda\lambda\,$}
\def\bv{\ifmmode{(B-V)}\else{$(B-V)$}\fi}

\def\lyalpha{Ly-$\alpha$}
\def\halpha{H$\alpha$}

\def\mgii{Mg$_2$}

\def\manq{\rm AGB-manqu\'e}

\def\magsec{\mathfont{{\rm mag\ arcsec}^{-2}}}
\def\mlam{\mathfont{m_{\lambda}}}
\def\Lvsun{\mathfont{{\rm L}_{V, \odot}}}
\def\Lbsun{\mathfont{{\rm L}_{B, \odot}}}

\title{Far-Ultraviolet Radiation From Elliptical Galaxies}

\author{Robert W. O'Connell}
\affil{Astronomy Department, University of Virginia, P.O. Box 3818,
    Charlottesville, VA 22903-0818}

\begin{abstract}

Far-ultraviolet radiation is a ubiquitous, if unanticipated, phenomenon
in elliptical galaxies and early-type spiral bulges.  It is the most
variable photometric feature associated with old stellar populations.
Recent observational and theoretical evidence shows that it is
produced mainly by low-mass, small-envelope, helium-burning stars in
extreme horizontal branch and subsequent phases of evolution.  These
are probably descendents of the dominant, metal rich population of the
galaxies.  Their lifetime UV outputs are remarkably sensitive to their
physical properties and hence to the age and the helium and metal
abundances of their parents.  UV spectra are therefore exceptionally
promising diagnostics of old stellar populations, although their
calibration requires a much improved understanding of giant branch
mass loss, helium enrichment, and atmospheric diffusion.

\end{abstract}


\section{INTRODUCTION}

Far-ultraviolet radiation was first detected from early-type galaxies
by the {\it Orbiting Astronomical Observatory-2} in 1969.  This was a
major surprise because it had been expected that such old stellar
populations would be entirely dark in the far-UV.  To the contrary,
not only did elliptical galaxies and the bulges of early-type spirals
contain bright UV sources, but their energy distributions actually
increased to shorter wavelengths over the range 2000 to 1200 \AA,
resembling the Rayleigh-Jeans tail of a hot thermal source with $\teff
\ga 20000$K.  The effect was therefore called the ``UV-upturn,'' the
``UV rising-branch,'' or, more simply, the ``UVX.''  It was only the
second new phenomenon (after X-rays from the active galaxy M87)
discovered by space astronomy outside our Galaxy.

Controversy flourished over the interpretation of the UVX for the next
20 years because of the slow accumulation of high quality UV data.
More recent evidence has winnowed the alternatives and strongly
supports the idea that the UVX is a stellar phenomenon (as opposed to
nuclear activity, for example) associated with the old,
dominant, metal-rich population of early-type galaxies.  It is the
most variable photometric feature of old stellar populations.  It
appears to be produced mainly by low-mass, helium-burning stars in
extreme (high temperature) horizontal branch and subsequent phases of
evolution.  Such objects have very thin envelopes ($M_{ENV}
\lesssim 0.05\, {\rm M}_{\odot}$) overlying their cores.
On both theoretical and observational grounds, the lifetime UV outputs
of these stars are exquisitely sensitive to their physical properties.
They depend strongly, for instance, on helium abundance; the UV spectrum
is the only observable in the integrated light of old populations with
the potential to constrain their He abundances.  More remarkably,
changes of only a few 0.01 \msun\ in the mean envelope mass of an
extreme horizontal branch population can significantly affect the UV
spectrum of an elliptical galaxy.

If this interpretation is correct, then far-UV observations become a
uniquely delicate probe of the star formation and chemical enrichment
histories of elliptical galaxies.  They do, that is, once we
understand the basic astrophysics of these advanced evolutionary
phases and their production by their parent populations.  However,
this is one of the last underexplored corners of normal stellar
evolution, and a complete interpretation is not yet at hand, even for
nearby systems such as globular clusters where full color-magnitude
diagram information is available.  The key physical process involved
in producing the small-envelope stars is mass loss during low-gravity
phases on the red giant branch and subsequent asymptotic giant
branch.  Serious modeling of mass loss has only recently begun, and
we so far have little intuition for the effects of population
characteristics such as metal abundance.  Although the interpretation
of the integrated light of galaxies has heretofore relied on
astrophysics established and tested in the context of local stars, it
may be that the UVX problem will be the first where observations of
galaxies will act as strong diagnostics of stellar evolution theory.
At any rate, it is clear that to understand the controlling mechanisms
of the UVX in galaxies we must conjoin integrated light observations
of distant galaxies with the stellar astrophysics of globular clusters
and hot field stars in our own and nearby galaxies.

There are broader ramifications of this interpretation as well.  UV
light acts as a tracer for stellar mass loss.  As the primary source
of fresh interstellar gas and dust in old populations, stellar mass
loss is directly linked to a diverse set of other important phenomena,
including gas recycling into young generations of stars, galactic
winds, X-ray cooling flows, far-infrared interstellar emission, dust
in galaxy cores, and gas-accretion fueling of nuclear black holes. The
UV light also traces the production of low-mass stellar remnants.  The
hot UVX stars, regardless of their origin, are important distributed
contributors to the interstellar ionizing radiation field of old
populations.  It is possible that the UVX is influenced by, and
therefore reflects, galaxy dynamics.  Finally,
characterization of the UV light of nearby ellipticals, its separation
into young or old stellar sources, and its predicted evolution is also
basic to the development of realistic ``K-corrections'' for
cosmological applications to high redshift galaxies and to
interpretation of the cosmic background light.

There has been excellent progress over the last decade in understanding
the UVX phenomenon, but the first question that might occur to the
reader is why it took 30 years simply to identify its source.  The
answer lies in our historically limited capability for extragalactic
UV observations, a subject we discuss in the next section.  Following
that, we describe the discovery of far-UV light from old populations
and its basic observational characteristics, the lively debate over
the leading alternative interpretations, and the confluence of theory
and new observations that has led to the currently accepted
interpretation.  We also discuss several of the other observational
opportunities presented by the generally faint UV background in
galaxies.  By ``early-type'' galaxies in this paper, we mean
ellipticals, S0s, and the large bulges of spirals of types Sa and Sb,
although most of the detailed analysis to date has concentrated on
Es and S0s.

\section{INSTRUMENTAL CONSIDERATIONS}

Progress in understanding the far-ultraviolet radiation from galaxies
has been more circumscribed by instrumental limitations than was the
case, for instance, in extragalactic X-ray astronomy.  Fewer
long-lived ultraviolet facilities have been available, and most of
these have not been well suited for the study of galaxies.  The
problems are both intrinsic and technical.  Intrinsically, galaxies
are faint, extended sources.  For typical elliptical galaxies,
incident far-UV photon rates per unit solid angle per unit wavelength
are typically over 50 times smaller than in the V-band.  The centers
of nearby bright ellipticals produce only a few $\times\, 10^{-15}\,
\flun {\rm arcsec}^{-2}$ at 1500 \AA\ averaged over a 10\arcsec\
radius (Burstein et al 1988,  Maoz et al
1996, Ohl et al 1998).  The paucity of high contrast spectral
features in UV hot star spectra at the spectral resolution and S/N
possible for E galaxies has also hampered interpretation.

There has never been a large area UV sky survey sensitive enough to detect
galaxies.  The only all-sky survey yet made in the UV was by TD-1 in
1973 (Boksenberg et al 1973).  This has a limit of about 9th magnitude and did not
include a single galaxy or QSO.  The GALEX mission (Martin et al
1997), now under development, will remedy this situation and produce a
survey up to 10 magnitudes fainter.  For now, however,
the fact remains that the deepest survey of the UV sky is comparable
to the Henry Draper catalog of stars, made around 1900.  So UV
astronomy, at least in this sense, is still 100 years behind optical
astronomy.

The technical development of UV instrumentation has been reviewed by
Boggess \& Wilson (1987, spectroscopy), O'Connell (1991, imaging),
Joseph (1995, detectors), and Brosch (1998, surveys).  UV telescopes
have been small, mostly less than 40 cm diameter.  Other than the
2.4-m Hubble Space Telescope (HST), the largest UV instrument
available has been the 1-m diameter {\it Astro} Hopkins Ultraviolet Telescope
(HUT), which as a Shuttle-attached payload had an equivalent dedicated
observing lifetime in 2 missions of only about 6 days (Kruk et al
1995).  Observations of galaxies are difficult with the
small entrance apertures available on most UV spectrometers,
for example the International Ultraviolet Explorer (IUE) ($10\arcsec
\, \times 20\arcsec$) or the HST/Faint Object Spectrograph ($\leq
1\arcsec$), which were designed for point sources.  With IUE, long
exposures of typically 4--8 hours were needed to register far-UV
spectra of galaxies.  Newer instruments are better matched to
requirements for galaxy work.  HUT was the first UV spectrometer
designed specifically for galaxies, with apertures as large as
$19\arcsec \times 197\arcsec$ (providing, however, only one spatial
resolution element).  The {\it Astro} Ultraviolet Imaging Telescope
(UIT) experiment, designed for filter imaging in the 1230--3200 \AA\
region, had a field of view (40\arcmin) and spatial resolution
(3\arcsec) well matched to ground-based studies of nearby galaxies.
The new Space Telescope Imaging Spectrograph (STIS) offers UV
apertures up to $2\arcsec \times 52\arcsec$, encompassing many spatial
resolution elements, and can image $25\arcsec \times 25\arcsec$ fields
with UV photon-counting detectors and 0.05\arcsec\ resolution.  The HST
Advanced Camera for Surveys, scheduled for installation in 2000, has
high throughput UV cameras with fields up to $30\arcsec \times
30\arcsec$.

The quantum efficiencies of UV detectors such as cesium iodide and
cesium telluride photocathodes are only modest (10--30\%), and net
throughputs are further compromised by the lower reflectivities and
transmissions of UV optical components.  The most widely used mirror
coating, magnesium fluoride, has a short-wavelength cutoff near 1150
\AA.  To obtain response to the Lyman discontinuity at 912
\AA\ special coatings such as silicon carbide are now available (e.g.\
Kruk et al 1995), though these do not achieve reflectances typical
of standard coatings at longer wavelengths.

Two special requirements for far-UV observations have serious
practical consequences.  First is the necessity to suppress the
effects of the strong geocoronal \lyalpha\ emission line at 1216 \AA.
This is usually straightforward in spectrographs, but in photometers
or imagers the only remedy is to use blocking filters that permit
response only for $\lambda \ga 1250$ \AA.  Second is the necessity to
suppress residual filter and detector response to long-wave ($\lambda
> 3000$ \AA) photons.  Even though this may be only a tiny fraction of
peak UV response, it covers a wide wavelength range.  Because cool
sources, such as stars with $\teff < 7000\,$K, can have optical
$f_{\lambda}$  thousands of times higher than their UV
$f_{\lambda}$, there can be serious ``red leak'' contamination of UV
observations.  Despite considerable effort (e.g.\ on Wood's filters),
it has not been possible to develop fully satisfactory long-wave
blocking devices with good peak UV response.  Therefore, red leak
suppression depends on the use of ``solar-blind'' detectors with large
photoelectron work functions, such as cesium iodide, which has very
small response for $\lambda > 1800\,$\AA.  Such detectors have been
used in most UV spectrometers but were not available in the HST Wide
Field Camera (WFPC2) or HST Faint Object Camera (FOC), both of which
consequently required careful red leak calibrations for use shortward
of 2500 \AA.  The effects of red leaks on HST photometry of stars and
galaxies can be dramatic and have been discussed by Yi et al (1995)
and Chiosi et al (1997).  The requirements for simultaneous
Ly-$\alpha$ and red leak suppression imply smaller bandwidths and
lower throughputs for far-UV imaging or photometry than is typical at
longer wavelengths. 

Because of these technical constraints, the working
``far-ultraviolet'' (FUV) band covers $\sim$1250--2000 \AA\ for
imaging or photometry, extended to about 1150 \AA\ for spectroscopy.
The ``mid-ultraviolet'' (MUV) band covers $\sim$2000--3200 \AA\ (3200
\AA\ being both the useful sensitivity limit of cesium telluride
photocathodes and the short-wavelength cutoff of the Earth's
atmosphere).  We will call the 3200--4000 \AA\ region accessible from
the Earth's surface the ``near-ultraviolet'' (NUV).  The 912--1150
\AA\ region in galaxies has been explored to date only by HUT,
though FUSE (scheduled for a 1999 launch) will also cover this range in
brighter objects.   

Unless noted, magnitudes quoted in this paper will be on the
monochromatic system, where $m_{\lambda} = -2.5 \log \rm{F}_{\lambda}
- 21.1 $ and $\rm{F}_{\lambda}$ is the mean incident flux in the
relevant band in units of 
\flun; the zero point is such that $m_{\lambda}(5500\, {\rm \AA}) =\,$
V.  Notation for colors will be, for instance, 1500--V $ \, \equiv\,
m_{\lambda}(1500\, {\rm \AA}) - \,$V.

\section{DISCOVERY AND ALTERNATIVE INTERPRETATIONS}

Prior to the first UV observations, there was a widespread expectation
that normal elliptical galaxies would be uninteresting in
the FUV (as, later, would also be the case with the X-ray and
far-infrared regions).  The hottest identified stellar component of
any consequence was the main sequence turnoff, with a temperature
($\teff \sim 6000\,$K) too cool to produce many FUV photons.  Although
it was recognized that the old, metal-poor populations of globular
clusters sometimes contained horizontal-branch (HB) stars with $\teff
\ga 10000\,$K, these were thought to be absent in the clusters (e.g.\
47 Tucan\ae) with metal abundances nearest those of massive galaxies.
The only hint of hot populations in E galaxies was the presence of
[O II] emission lines, though these could plausibly be explained
without stellar photoionization (Minkowski \& Osterbrock 1959).

\begin{figure}
\plotfiddle{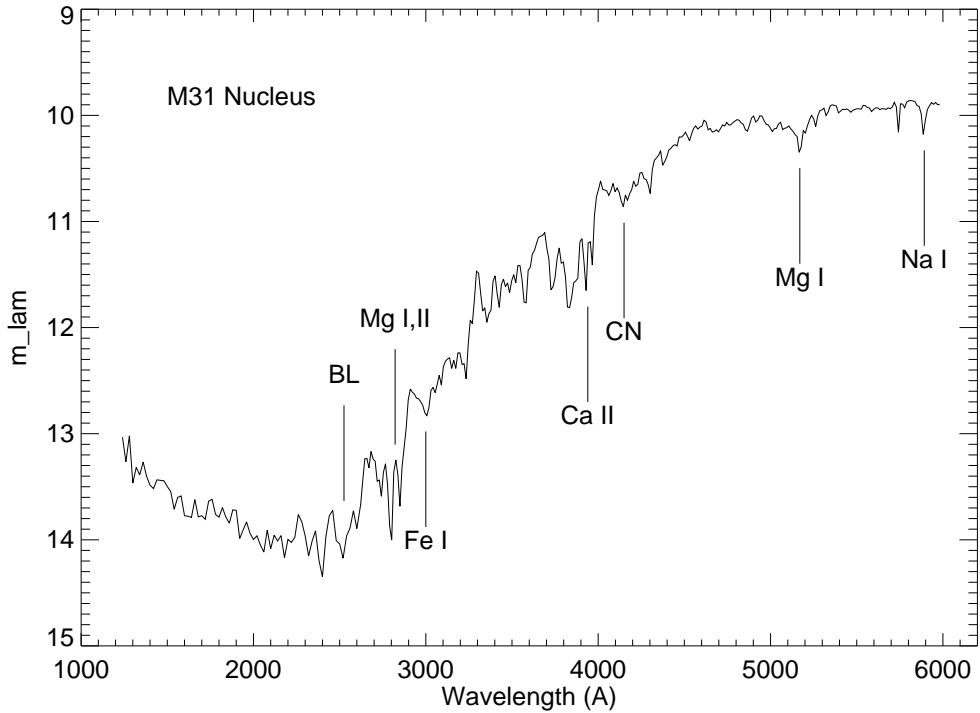}{3.5truein}{0}{80}{80}{-248}{-300}
\caption{
A composite UV-optical energy distribution
for the center of the Sb galaxy M31.  IUE data taken with a $10\arcsec
\times 20\arcsec$ aperture is plotted below 3200 \AA, while a
ground-based spectrum covering the same region is plotted above.
Resolution is 20 \AA\ below 2600 \AA\ and 12 \AA\ above.
Irregularities in the UV spectrum below 2200 \AA\ are mainly noise.
Some of the stronger absorption line features are identified (``BL''
corresponds to a strong blend of Fe and other metallic lines near 2538
\AA).  The ``UV-upturn'' is the rise in the spectrum at wavelengths
shorter than 2000 \AA.  By simple extrapolation of the far-UV
continuum slope, one finds that the upturn component contributes only
about 0.3\% of the V light of the galaxy.  Spectrum courtesy of D
Calzetti. 
}
\end{figure}

FUV radiation from galaxies was first detected by the University of
Wisconsin UV photometer carried on the second {\it Orbiting
Astronomical Observatory} (OAO-2).  The experiment obtained fluxes with
an entrance aperture of $10\arcmin$ diameter in 7 intermediate band
filters extending from 4250 \AA\ to the FUV at 1550
\AA.  The first announcement (Code 1969) of results for an old
population was for the central bulge ($r < 900\,$pc) of the Local Group
Sb spiral M31.  As expected, the energy distribution of M31 fell
steeply between 3500 and 2500 \AA\ but then, remarkably, began to rise
again at shorter wavelengths.  A more recent UV-optical spectrum of M31
is shown in Figure 1.  Since the energy distributions of normal stars
cooler than $\teff \sim 8500\,$K (spectral type A5) decline
precipitously below 1800~\AA\ owing to absorption by metallic
ionization edges (e.g.\ Fanelli et al 1992), the detection of any
far-UV flux in galaxies implies sources with higher equivalent
temperatures.  After a difficult calibration process, OAO-2 photometry
was ultimately published for 7 E/S0 objects and the M31 bulge (Code et
al 1972, Code \& Welch 1982).  The OAO-2 detections of two objects
were confirmed, and new detections made of another 11 E galaxies, by
the {\it Astronomical Netherlands Satellite} (launched in 1974) using
intermediate band photometry with a $2.5\arcmin \times 2.5\arcmin$
aperture over the range 1550--3300 \AA\ (de Boer 1982).

An immediate conclusion from the UVX observations which was emphasized
by Code and his colleagues was that early-type galaxies exhibited much
larger scatter in the UV than was expected from their conspicuously
homogeneous behavior in the optical to near-IR (4000--20000 \AA)
region.  The UV observations implied divergent histories at some level
and were among the first indications that elliptical galaxy
populations were more heterogeneous than envisioned in Baade's classic
definition of Population II (Baade 1944, O'Connell 1958, O'Connell
1980, Faber et al 1995).  They called into question the use of E galaxies as
``standard candles'' in cosmological studies.  They also complicated
the construction of accurate K-corrections needed to transform
photometry of high redshift elliptical galaxies to standard bands in
the restframe (e.g.\ Pence 1976, Coleman et al 1980, King
\& Ellis 1985, Bertola et al 1982, Kinney et al 1996).  

Interpretation of the unexpected OAO-2 results was initially confused
by calibration uncertainties which produced anomalously steep FUV
energy distributions (in normal spirals and irregulars as well as
early-type systems, see Code \& Welch 1982).  Code (1969) and Code et
al (1972) suggested that the UVX component was nonthermal
radiation from an active nucleus (AGN) or scattering of photons from
massive hot stars by interstellar dust.  The latter would have implied
that most E/S0 galaxies contain an appreciable Population I
component.

Hills (1971) pointed out that the steep rise of the M31 UV spectrum to
higher photon energies was incompatible with known nonthermal sources
but was closely matched by the Rayleigh-Jeans tail of a high
temperature thermal source.  Based on comparison with a
small sample of UV-bright stars in the globular cluster M3, he
proposed that the UV upturn is produced by highly evolved, hot,
low-mass stars such as the central stars of planetary nebulae, now
known as post-asymptotic giant branch (PAGB) stars, or their hot white
dwarf descendents.  He did not require that these be members of a
strong Population II (old, metal-poor) component but pointed out that
their prominence would probably depend on metal abundance.

Tinsley (1972a) argued that the UV light arose instead from young,
massive, main sequence stars and showed that a spectral synthesis
model for an old galaxy with an exponentially declining star formation
rate and an e-folding time of 2 Gyr could fit the OAO-2 flux for M31
observed at 1700 \AA.  This would imply that the UVX was related to a
normal, if temporally extended, star formation process in early-type
systems.  Fuel for the star formation might be primordial gas consumed
gradually over a Hubble time, mass loss from red giants, or material
accreted from outside galaxies (Gallagher 1972, Tinsley 1972b,
O'Connell 1980, Gunn et al 1981).  Residual star formation histories
of the type suggested by Tinsley would drastically change the
predicted properties of E galaxies viewed at moderate look-back times,
whereas the low-mass star interpretation would have less serious
implications for spectral evolution.

It was implicit in these early studies that the hot components that
dominated the far-UV light could be virtually undetectable at visible
wavelengths---i.e.\ that the UV was providing entirely independent
information about galaxies.  Ignoring any contribution from the cool
components to the UV light, the maximal fractional
contribution of a hot component to the integrated V-band light of a
galaxy will be $p_{\it max} \sim 10^{0.4\,\Delta}$, where $\Delta = $
(1500--V)$_{\rm hot} -$ (1500--V)$_{\rm obs}$.   A color for a typical
E galaxy is (1500--V)$_{\rm obs} \sim +3$ while a component with
an appropriate far-UV spectral slope (B0
equivalent) has (1500--V)$_{\rm hot} \sim -4.5$, implying that
$p_{\it max} \sim 0.001$.  This is about 50 times smaller than
could be directly detected in the V-band using spectral synthesis
techniques.

The early workers on the UVX realized that the best tests of the
alternative interpretations were {\it (a)} UV spatial structure and
{\it (b)} UV spectral features observed at higher resolution.  An
active nucleus would be a concentrated point source, whereas a
population of low-mass stars would presumably have a smooth
distribution similar to that found in the optical bands for bulges and
E galaxies.  Young, massive-star populations would likely have a
clumpy structure, similar to the OB associations found in spiral arms,
and they might well be concentrated to disks.  In nearer galaxies
individual massive OB stars could be isolated.  Spectroscopically, a
UV-bright AGN would be easy to identify on the basis of broad,
high-excitation emission lines.  Active massive star-forming regions
characteristically exhibit strong UV resonance lines of Si IV, C IV,
and other species, often with P-Cygni profiles (e.g.\ Kinney et al
1993), whereas the spectra of hot, low-mass stars are relatively
weak-lined in the 1200--2000 \AA\ region.

Because of limited UV observing opportunities, it would not be
possible to apply these tests in a definitive way until over a decade
after the discovery of the UVX.  Only short-duration sounding rocket
or balloon experiments were available until 1978.  The most productive
observing facility for the study of the UVX in the period 1978--1990
was IUE (Kondo 1988).  IUE's handicaps of small
effective collecting area, small entrance aperture, and limited
dynamic range were beautifully compensated by its record 18 year
lifetime and a capability for very long integration times, and it
produced an invaluable set of UVX spectra.  The fact that its point
spread function was smaller than its $10\arcsec \times 20\arcsec$
entrance aperture also meant that spatial structure could be studied
to a radius of 10\arcsec.  After 1990, HST and the two {\it Astro}
missions provided new capabilities to study the UVX.  

In the next two sections, we describe the basic phenomenology of far-UV
sources in bright early-type galaxies, as determined by IUE and other
instruments, and how this bears on the now accepted interpretation of
these as low-mass stars in old stellar populations.

\section{SPATIAL STRUCTURE OF THE UVX}

\subsection{\it Evidence Against Young Stars}

UV imaging of early-type galaxies began in the 1970s.  Early rocket and
balloon experiments obtained low S/N images of the central bulge of
M31 which showed that it was an extended source in the far-UV with $r
\ga 4\arcmin$ (Deharveng et al 1976; Carruthers et al 1978;
Deharveng et al 1980).  This was sufficient to exclude the AGN
interpretation (in this particular case) but could not readily
distinguish between the old and young star models or other types of
diffuse sources.  A later rocket imaging experiment by Bohlin et al
(1985) provided far- and mid-UV photometry of M31 with 20\arcsec\
resolution.  The UV intensity profiles of the bulge were smooth and
similar to Kent's (1983) R-band profile for $r \la 1.1\arcmin$.
Although localized regions of massive star formation were readily
detectable in the outer spiral arms, similar structures were absent in
the bulge, nor could individual bright OB stars be detected there.

IUE observations of bright early-type galaxies (including M31, M32,
NGC 3379, NGC 4472, M87, NGC 4552, and NGC 4649) confirmed the spatial
extension of the far-UV light, even in the case of the prominent AGN
of M87, and indicated that it paralleled the profile of the visible
light, at least over the innermost 10\arcsec\ (Bertola et al 1980,
Perola \& Tarenghi 1980, N\o rgaard-Nielsen \& Kj\ae rgaard 1981, Oke
et al 1981, Bertola et al 1982, O'Connell et al 1986).
Deharveng et al (1982) and Welch (1982) used multiple IUE spectra to
study the light distribution within the inner 15\arcsec\ of the M31
bulge.  They obtained a smooth profile, unlike those of star-forming
regions, but found that the FUV light was slightly more concentrated
to small radii within this region than MUV or B band light, producing
gradients of several 0.1 mags in colors. The smooth distribution of
UVX light in these cases and its similarity to the optical band
profile, where old stars dominate, strongly suggested that old
stars produced the UVX.

In another rocket experiment, Onaka et al (1989) and Kodaira et al
(1990) obtained low-resolution, wide-field UV images of the Virgo
cluster, extending the earlier photometry of Smith \& Cornett (1982)
at 2400 \AA\ to 1600 \AA.  By comparing their total fluxes with IUE
values for the nuclei, Kodaira and colleagues found evidence of large UV
color gradients in five Virgo E galaxies.  They attributed the blue
nuclear excesses and the observed scatter in 1500--V colors to recent
star formation from galactic cooling flows, though their observations
were also consistent with gradients in low-mass populations.

\vskip 1.7truein
\centerline{[FIGURE 2: see 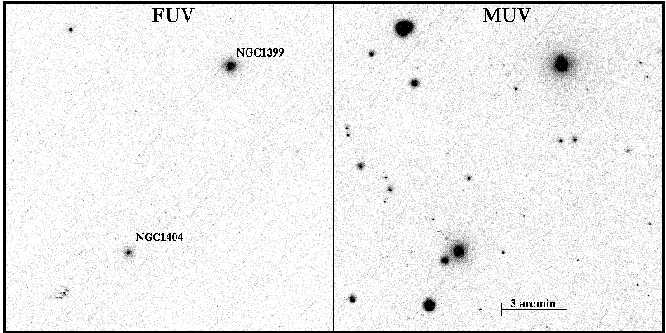]}
\vskip 1.7truein
\noindent Fig. 2 --- {\it Astro}/UIT images of the Fornax cluster
ellipticals NGC 1399 and 1404 in broad bands in the far-UV (1500 \AA)
and mid-UV (2500 \AA) with spatial resolution of $\sim 3\arcsec$.  The
mid-UV band is dominated by the main sequence turnoff.  All of the
far-UV light is from the UVX component.  It is smooth, without
evidence for massive stars, though is more concentrated than the
mid-UV light.  NGC 1399 has one of the strongest UVX components yet
discovered.  Note that the foreground stars have mostly vanished in the
FUV band; this is a pictorial representation of how unusual are the
objects which make up the UVX.

\vskip 0.1truein

\setcounter{figure}{2}

The best available set of large area UV maps of early-type galaxies
was obtained by the {\it Astro} UIT experiment during two Space
Shuttle missions in 1990 and 1995 (Stecher et al 1997).  Twenty-two
ellipticals and early-type (S0-Sb) spiral bulges were imaged with good
S/N at 3\arcsec\ resolution, and results for 10 of these have been
published (O'Connell et al 1992, Ohl et al 1998).  UIT images of
two Fornax cluster elliptical galaxies are shown in Figure 2.  In the
best cases, it was possible to obtain UV surface brightness profiles to
$\mu_{\lambda}
\sim 27\, \magsec$.  All of the objects exhibit smooth UV profiles
(except M87, in which the nonthermal jet is bright), with none of the
clumpiness normally associated with recent massive-star formation, and
the FUV contours are consistent in shape and orientation with
optical-band isophotes.  There is little evidence for dust lanes or
clouds in the galaxy centers; such features should be readily
detectable because of the high selective UV extinction of normal
dust.  In the M31 bulge, the point source detection threshold was
$\mlam$(2500 \AA) $\sim 18.4$, which excluded the presence of
individual main sequence stars hotter than B1 V (O'Connell et al
1992).  Over 200 such objects would be expected in the central
2\arcmin\ of the bulge if massive stars formed with a normal initial
mass function produced the FUV light.  The FUV profiles of about half
the sample are well fitted by de~Vaucouleurs functions ($\mu \sim a
+br^{0.25}$), which are characteristic of spheroids at optical
wavelengths.  However, the inner FUV profiles of several objects (NGC
3379, 4472, and 4649) are more consistent with an exponential function
(Ohl et al 1998).  Although exponentials are normally associated
with disks, the FUV isophotal contours are congruent to the B-band
contours, and the 3-dimensional FUV light distributions are therefore
unlikely to be genuinely disklike.  (Because of the large UV/optical
color gradients discussed below, it is not necessarily expected that
the UV profiles of objects that are true spheroids at optical
wavelengths would be closely de~Vaucouleurs in shape.)

High-resolution UV imaging from HST (mainly of smaller $\leq 22\arcsec
$ nuclear fields with the FOC) has confirmed the absence of massive
stars in the centers of M31 and M32 (King et al 1992, Bertola et al
1995, King et al 1995, Cole et al 1998, Brown et al 1998a, Lauer et al
1998) and in most UVX sources in the nuclei of 56 early-type galaxies
in the 2300 \AA\ survey of Maoz et al (1996).

The collective evidence of all these structural studies is that the
far-UV light in most early-type galaxies originates in a stellar
component with dynamics characteristic of the bulk of the old stellar
population.  Active nuclei or young massive stars are not important in
most cases.

\subsection{\it Structural Variations}

However, the UV-bright population is not a simple extension of the
well-studied, optically bright one.  This was evident from the scatter
in the ratio of UV to optical light first reported by Code et al
(1972) and Code \& Welch (1982), which has been amply confirmed by
later observations (see \S 5.2).   In addition, large internal
gradients in UV/optical colors have been revealed in almost all cases
studied with sufficient S/N.  Five of the galaxies shown in Figure 3
from the Ohl et al (1998) UIT sample display large internal 1500--B color
gradients with net changes up to $\sim 1.0$ mag over the region
photometered.  The 1500--B colors of 7 of the 8 objects become redder
outward, meaning that the far-UV light is more concentrated to the
galaxy centers than the optical light.  Both in amplitude and sign,
these changes are dramatically unlike the very mild, bluer-outward
color gradients encountered in the optical and IR (e.g.\ Peletier et
al 1990).  M32 is the only object which becomes bluer in 1500--B at
larger radii.

\begin{figure}
\plotfiddle{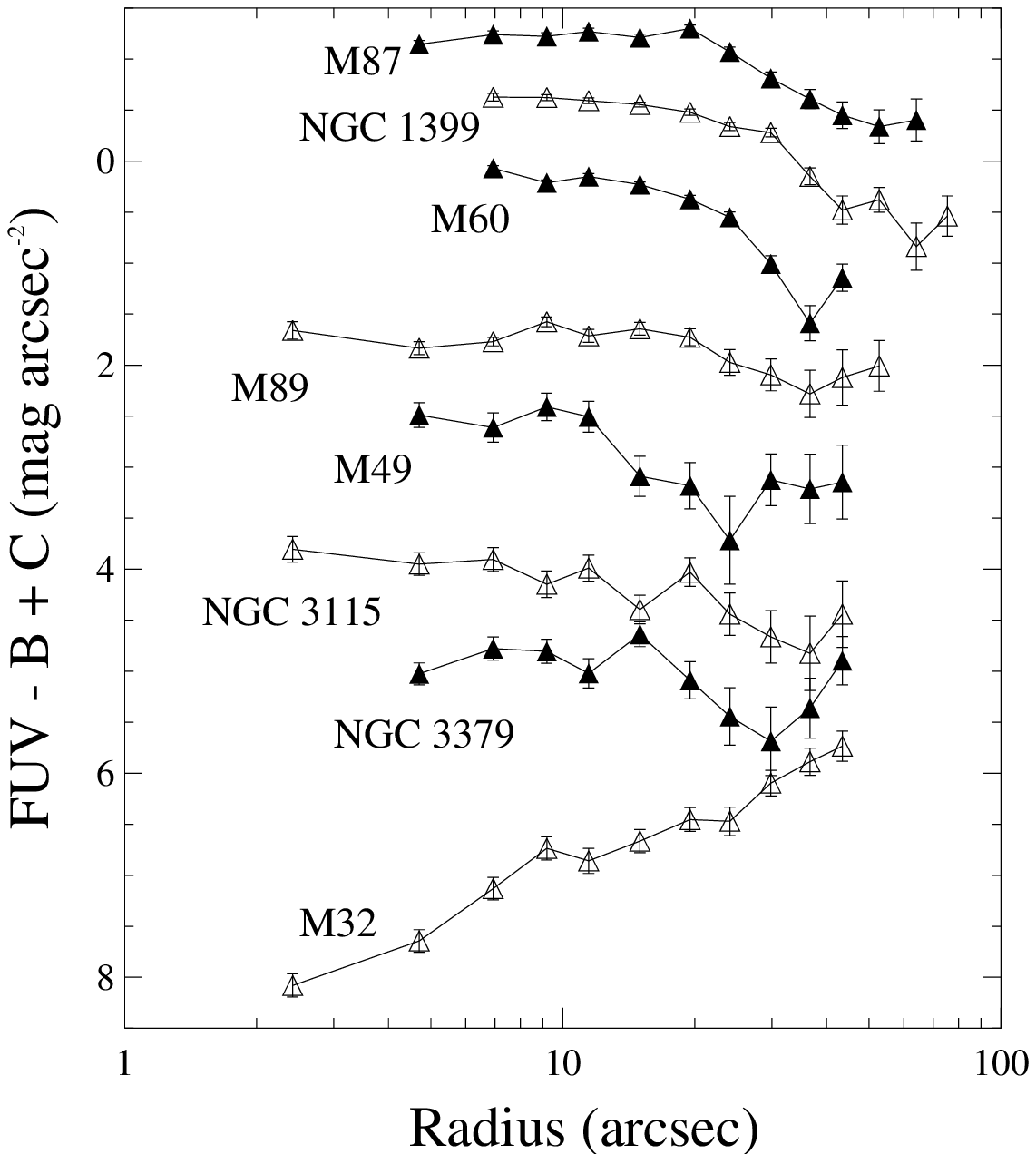}{4.5truein}{0}{90}{90}{-180}{-10}
\caption{
Radial FUV--B color profiles for 8 early-type
galaxies obtained by comparing {\it Astro}/UIT far-UV surface
photometry with B-band data from the literature.  The curves have been
offset for clarity and arranged in order of increasing central UVX.
One sigma error bars are shown. FUV--B colors redden with increasing
radius in all cases except M32, which shows a strong, reversed
profile.  It is the only object currently known to have this
behavior.  There is an interesting two-component structure in most of
the profiles.  Offsets in order from the top down are $\mathrm{C =
-2.5}$, $\mathrm{-2.0}$, $\mathrm{-1.5}$, $\mathrm{0.0}$,
$\mathrm{0.0}$, $\mathrm{+1.0}$, $\mathrm{+2.0}$, and $\mathrm{+3.5}$
mag $\mathrm{arcsec^{-2}}$.  From Ohl et al (1998).
}
\end{figure}

Internal extinction by dust cannot be responsible for these
gradients.  Aside from the absence of dust structures in the images
and the sense of the typical gradient (implying more extinction at
larger radii), the gradients are so large that significant
optical-band effects would be expected, since $A(4400{\rm \AA}) \sim
0.5 A(1500$\AA), where $A$ is the total extinction in magnitudes.  HUT
spectroscopy also places strict limits on the amount of
internal extinction (Ferguson \& Davidsen 1993, Brown et al 1997).
Instead, the gradients are apparently driven by a radial change in the
properties of the old star population.

\section{SPECTRAL AND PHOTOMETRIC CHARACTERISTICS OF THE UVX}

\subsection{\it Incidence, Spectral Shape, and Line Features}

Except in cases of obscuration by a major dust lane (e.g.\ in
edge-on S0-Sb objects), far-UV radiation has been detected in all
nearby early-type systems observed with adequate S/N.  As noted above
(\S 3), this implies the presence of sources with $\teff > 8500\,$K.
Rifatto and colleagues (1995a, 1995b) have compiled all UV
observations of galaxies published before 1990 and attempted to place
them on a homogeneous system, which is a challenge owing to the
varied types of experiments involved and the relatively
low photometric precision which is typical.  Their list includes 94
galaxies of type Sb or earlier with UV detections at $\lambda <
2100\,$\AA.  The list does not include later photometry or imaging
from the SCAP/FOCA balloon experiments (Milliard et al 1992, Donas et
al 1995, Treyer et al 1998), {\it Atlas}/FAUST (Deharveng et al
1994), {\it Astro}/UIT (Stecher et al 1997), {\it Astro}/HUT
(Kruk et al 1995), or the Maoz et al (1996) HST/FOC nuclear survey.
Combined, these roughly double the total number of far-UV E-Sb detections,
and HST is continually enlarging this sample.  It is worth emphasizing
that extragalactic UV observations are largely confined to relatively
nearby, bright systems (except for very distant objects where the
redshift brings the restframe UV into the bands accessible from the ground).

The early IUE spectra of the nuclei of bright ellipticals and spiral
bulges (Johnson 1979, Bertola et al 1980, Perola \& Tarenghi 1980,
N\o rgaard-Nielsen \& Kj\ae rgaard 1981, Oke et al 1981, Bertola et
al 1982, Deharveng et al 1982, O'Connell et al 1986) showed
immediately that the strong, broad emission lines characteristic of
active nuclei were absent, excluding the AGN hypothesis.  Signals for
$\lambda \la 2400\,$\AA\ were, however, very weak and subject to
several kinds of detector noise (which generated some spurious claims
of narrow coronal or chromospheric emission lines).  Except in the
brightest sources, it was necessary to average far-UV fluxes over
bandwidths of $\sim 50\,$\AA.

\begin{figure}
\plotfiddle{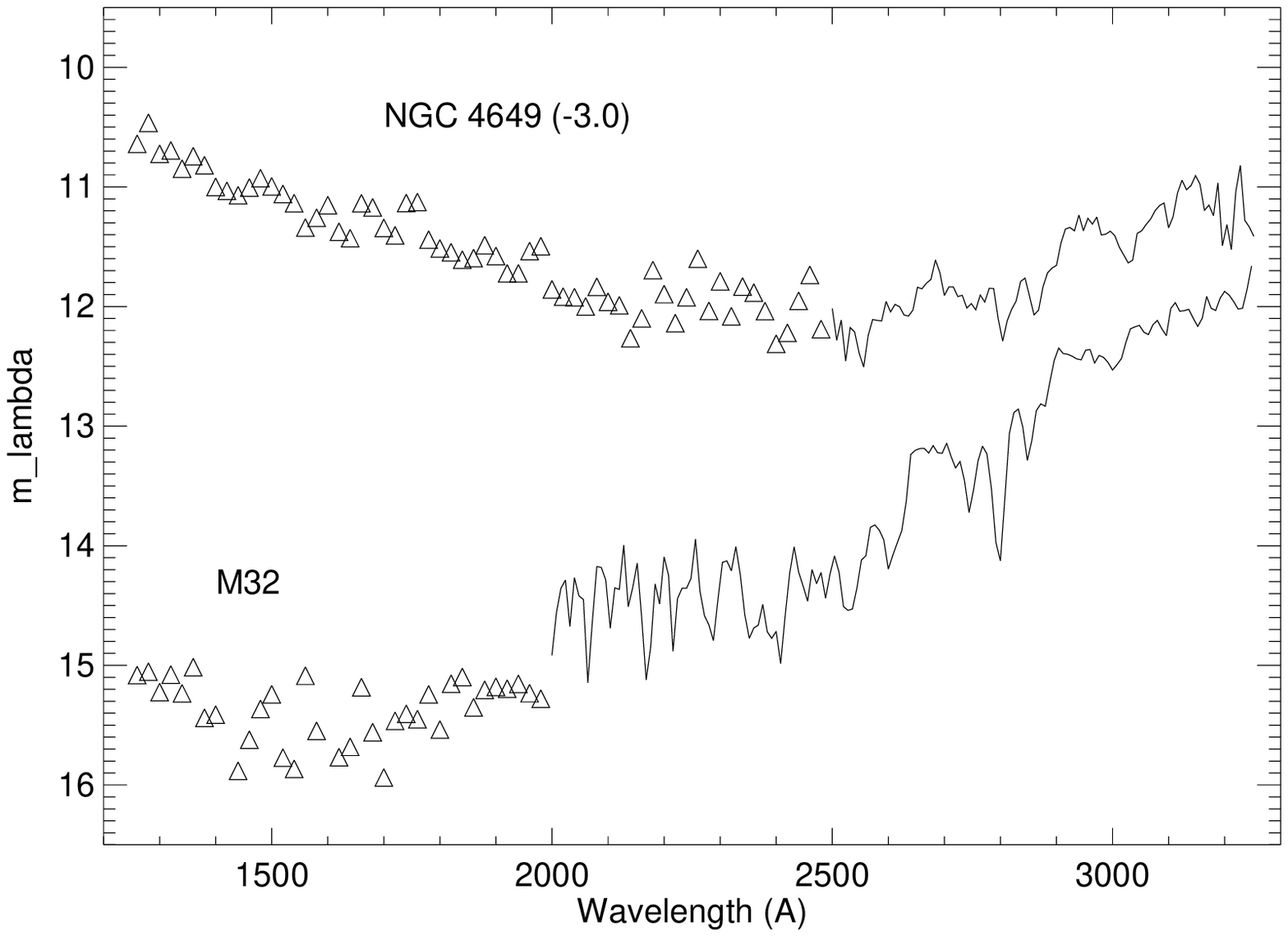}{3.5truein}{0}{80}{80}{-260}{-300}
\caption{
IUE spectra of two galaxies lying at the
extremes of UVX behavior.  M32 has the smallest known UV upturn, while
NGC 4649 has one of the strongest.  Data plotted with open triangles
have 20 \AA\ binning, while the solid line has 8 \AA\ binning.  The zero
point of the NGC 4649 spectrum has been shifted $-$3.0 mags. Neither
spectrum is corrected for redshift or foreground extinction.  The FUV
data for M32 are too poor to judge the slope of its UVX component.  The
weakness of the absorption lines near 2800 \AA\ in NGC 4549 is
caused by filling by the smooth UVX component, which contributes
over 70\% of the light at 2700 \AA\ in this object.  Reprocessed and
recalibrated IUE spectra courtesy of RC Bohlin.
}
\end{figure}

Burstein et al (1988, hereafter BBBFL) produced the largest
homogeneous set of good IUE spectra for early-type galaxies (32
objects).  Two examples are shown in Figure 4.  The flux rapidly
declines shortward of 3300 \AA.  Strong absorption features from Mg I
(2852 \AA), Mg II (doublet at 2800 \AA), Fe I (numerous lines), and
other metallic species are easily detectable down to $\lambda \sim
2500\,$\AA, as are strong discontinuities caused by metallic
blanketing at 2640 and 2900 \AA.  The mid-UV line spectrum closely
resembles that of F-G dwarf stars (see Fanelli et al 1992), the
spectral types expected for the main sequence turnoff in an old
E-galaxy population.  The flux reaches a minimum in the range
2000--2600 \AA, where the S/N is almost always rather poor, then
usually rises steeply again to shorter wavelengths.  No maximum is
detected in the rising component longward of the IUE cutoff at $\sim
1150\,$\AA.  

At the resolution permitted by the noise, the typical far-UV IUE
spectrum is a relatively smooth continuum with an equivalent
temperature  $\teff \ga 20000\,$K.  The spectral slope of the
upturn is roughly constant, so that the far-UV rise begins at longer
wavelengths in galaxies with brighter UVX components (e.g.\ NGC 4649
in Figure 4).  The contribution of the UVX component to the mid-UV light
can be appreciable, ranging up to 75\% at 2700 \AA\ for objects like
NGC 4649 (BBBFL, Ponder et al 1998, Dorman et al 1999), though
this drops rapidly at longer wavelengths because of the steep rise in
the spectrum of the cooler main sequence turnoff stars.  In objects
with the smallest UVX components (e.g.\ M32 and NGC 4382), the
spectra appear to flatten below 2000 \AA, rather than rise, but the
S/N is too poor to estimate a temperature (BBBFL and Figure 4). 

In the great majority of cases, E galaxy spectra longward of 3200 \AA\
are quite similar to one another; the large spectral anomalies associated
with the UVX are confined to the vacuum UV.  This suggests that
the stars responsible for the UVX are well segregated in the
color-magnitude diagram from the bulk of the population.  An
interesting exception is M87, where anomalies are detectable up to
4000 \AA\ (Bertola et al 1982, McNamara \& O'Connell 1989); they are
spatially extended and may be related to massive star formation in
M87's cooling flow (see \S 8).

As originally emphasized by Tinsley (1972a), the characteristic FUV
slope of UVX galaxies is consistent with star-forming models in which
massive O and early B stars dominate the light.  These require star
formation to have occurred within the last 10--20 Myr.  Tinsley did not
discuss the far-UV spectral shape of her models, but later studies
(e.g.\ Wu et al 1980, Gunn et al 1981, Nesci \& Perola 1985,
Rocca-Volmerange \& Guiderdoni 1987, Bica \& Alloin 1988, Burstein et
al 1988, Ferguson et al 1991, Bruzual \& Charlot 1993) would show that
the young star and old, evolved star models could produce almost
indistinguishable FUV energy distributions at low spectral resolution
(see Figure 5).  It would be necessary to consider other information,
especially spectral features, to resolve the ambiguity.

Because of the limited signal-to-noise of individual spectra, IUE
studies of possible far-UV absorption lines have been based on
summed spectra for either the same or several objects.  A serious
complication is a systematic background of ``fixed pattern'' and camera
artifact features in long-exposure IUE spectra 
(Crenshaw et al 1990).  Welch (1982) analyzed 12 exposures of
the center of M31 and detected weak absorption features at 1260 \AA\
(Si II $+$ S II), 1302 \AA\ (O I $+$ Si II $+$ Si III), and 1335 \AA\
(C II).  The features were confirmed by BBBFL in M31, but they were
weak or absent in a summed spectrum for 3 bright UVX E galaxies
(BBBFL).  These lines are characteristic of normal early-B stars (e.g.
Fanelli et al 1992) but are considerably stronger in the stars than
in any of the UVX sources.  The strong Si IV (1400 \AA) or C IV (1550
\AA) absorption features associated with massive O stars were absent
in both M31 and the summed E spectrum.

Welch (1982) and BBBFL argued that the weakness of the massive OB-star
spectral features was inconsistent with recent star formation as the
source of the UVX.   BBBFL pointed out additional evidence in the
form of continuum shapes.  Although it is possible to produce a
star-forming model whose spectrum matches the typical steep UVX far-UV
spectral slope (see Figure 5), in fact many systems with younger
populations (e.g.\ NGC 205 or 5102) have rather flat energy
distributions in the 1200--3000 \AA\ region.  This is the signature of
an aging starburst or a young starburst containing local extinction
(Kinney et al 1993).  Furthermore, the predicted young-star
contamination of the optical band if the UVX originates in
massive stars is significantly larger than limits from careful
spectral synthesis studies----e.g.\ the 2\% maximum at 4000 \AA\ set
by Rose (1985) using high-resolution spectra of 12 E galaxies.  The
uniformity of the UV-upturn slope and the absence of warm-star effects
at wavelengths longer than 2000 \AA\ are therefore additional evidence
against the involvement of massive stars in the UVX phenomenon.  

\begin{figure}
\plotfiddle{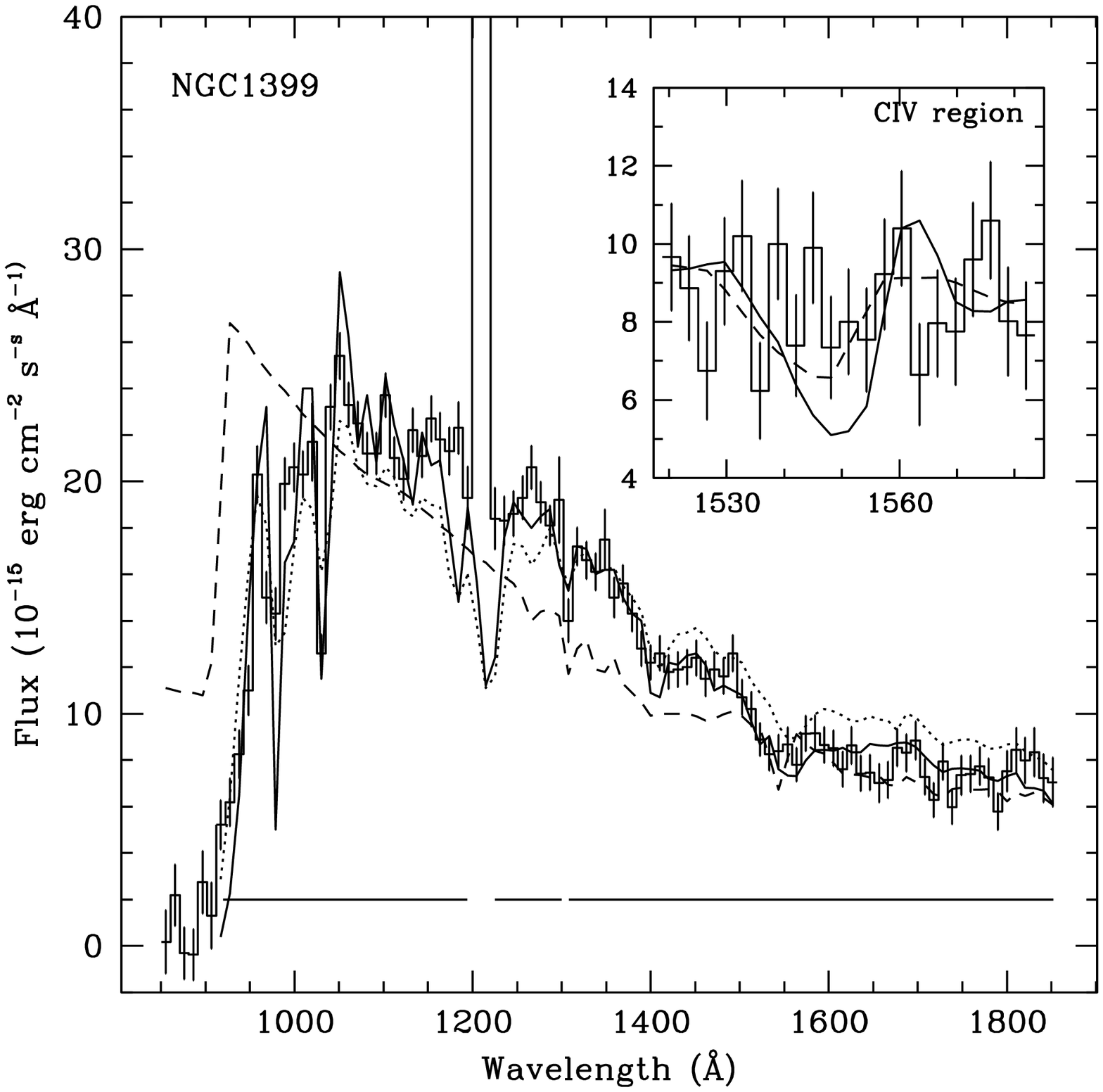}{4.0truein}{0}{60}{60}{-185}{-100}
\caption{
 The {\it Astro}/HUT spectrum of the gE galaxy
 NGC 1399 in the Fornax Cluster (see Figure 2).  The histogram is the
 observed flux in 10 \AA\ bins.  The solid line shows the best-fitting
 Kurucz (1991) solar abundance model atmosphere, which has $\teff =
 24000\,$K. The dashed line is a model from Rocca-Volmerange \& Guiderdoni
 (1988) for an old galaxy with continuing star formation.  This contains
 hotter starlight than is present in the galaxy.  The inset shows the
 observed spectrum near the C IV 1550 \AA\ doublet compared with
 continuous star forming models.  From Ferguson et al (1991).
 }
\end{figure}

The best far-UV spectra of UVX galaxies, covering the range 900--1800
\AA, were obtained by {\it Astro}/HUT with a photon-counting detector
and calibration superior to IUE's.  HUT spectra of 8 early-type
systems confirmed the weakness of the massive OB-star spectral
features (Ferguson et al 1991, Davidsen \& Ferguson 1992, Brown et
al 1995, Brown et al 1997).  For example, they placed an upper limit on
the strength of C IV 1550 \AA\ of $\la 3\,$\AA\ equivalent width in
NGC 1399, which formally excluded spectral synthesis models for
continuous star formation with normal metal abundances (see Figure 5).  Equally
important, by extending spectral coverage below 1150 \AA, HUT was
able to detect turnovers in the UVX spectra that place firm upper
limits on their effective temperatures of $\teff \la 25000\,$K
(equivalent to a B0 V star).  This excludes continuing star formation
models with a normal initial mass function (IMF).  Only contrived
models (e.g.\ invoking a truncated IMF or an unprecedented
synchronization of star formation in different galaxies) can reconcile
young populations with these results.  The narrow $\teff$ range, however,
also appears to require an unusual degree of ``fine-tuning'' in an old
population interpretation.

The HUT spectra also provide excellent limits on the amount of
internal interstellar extinction.  Because the slope of the far-UV
spectral rise is nearly at the maximum encountered among hot stars
(Dean \& Bruhweiler 1985, Fanelli et al 1992), there is little room
for interstellar reddening.  In the HUT data, there is no evidence for
extinction significantly in excess of the expected Galactic foreground.
Since dust is normally associated with star-forming regions, this is
yet further evidence of their unimportance in UVX galaxies.  

The spectroscopic evidence therefore strongly corroborates the 
structural evidence from the last section that massive stars
are not responsible for the UV-upturn.

Recently, Bica et al (1996) have produced composite spectra for
groups of E galaxies from the IUE archives.  They find evidence for
broad absorption features at 1400 and 1600 \AA\ in most UVX sources,
which they identify with the Ly-$\alpha$ satellite lines in
intermediate-temperature (DA5) white dwarf stars.  If true, this would
be remarkable since only a very unusual population would have the
requisite concentration of white dwarfs.  The features are not present
in the better quality HUT spectra discussed above, but Bica et al
suggest they are confined only to the nuclei and have been diluted by
the larger HUT entrance aperture.  This can be checked with HST/STIS.

\subsection{\it Amplitude and Correlations with Other Properties}

The most striking feature of the UVX phenomenon is its large
variation from object to object.  As measured by the color 1500--V,
the amplitude of the UVX in the centers of bright E-Sb galaxies varies
from $\sim 4.5$ to $\sim 2.0$, which is a factor of 10 in the ratio
of far-UV to visible flux.  No optical-IR photometric or spectral
index for normal old stellar populations exhibits a comparable range;
in fact, most do not vary more than $\pm$30\% (e.g.\ Sandage \&
Visvanathan 1978, Peletier et al 1990, Trager et al 1998).  The
UV variations are not confined to the nuclear regions observed with IUE;
they are also present in all the large aperture data sets cited above
(OAO, ANS, UIT, HUT, FAUST, and the various rocket/balloon
experiments).

Large excursions of this kind are usually associated with an
incidental component rather than with the aggregate population of a
galaxy.  To what extent does the UVX convey useful information about
the fundamental properties of galaxies?

A key insight was provided by Faber and her colleagues (Faber 1983,
BBBFL):  the UVX appears to be stronger in more metal-rich galaxies.
Faber (1983) found in a small sample of early-type galaxies that
nuclear UV colors became bluer as the nuclear spectral line index
Mg$_2$, which measures the Mg~I $+$ MgH absorption features near 5170
\AA, increased.  The Mg features are produced by the dominant old
stellar population, and this correlation simultaneously links the UVX
to the bulk of the galaxy while further weakening the case for massive
stars (since there is no obvious reason why recent star formation
would be related to metal abundance).  Interestingly, the correlation
is reversed in sense from the well-known dependences of (U--B) or
(B--V) colors on metal abundance in old populations.  Driven
mainly by opacity effects in stellar envelopes and atmospheres, these
become redder as abundance increases.

\begin{figure}
\plotfiddle{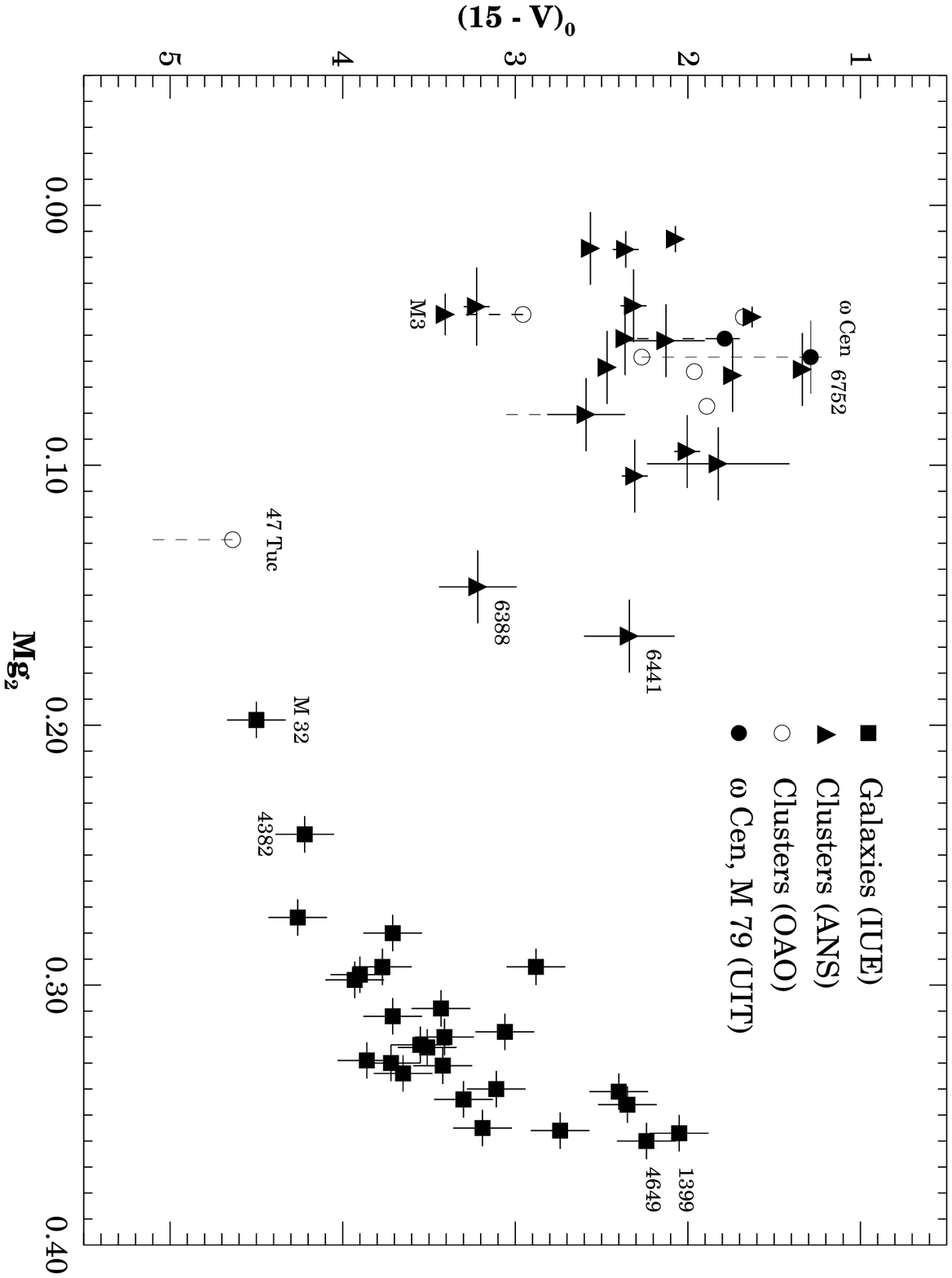}{4.5truein}{90}{70}{70}{+285}{-40}
\caption{
Amplitude of the UVX in old stellar
populations, as measured by the color 1500--V, as a function of the
\mgii\ line index, which measures absorption from Mg I $+$ MgH near
5170 \AA.  The E galaxy data are from IUE, mostly from the study by
BBBFL.  The globular cluster data are from several sources, as indicated
in the legend.  The clusters and galaxies are clearly distinct kinds of
populations.   From Dorman et al (1995).  
}
\end{figure}

The correlation was confirmed by the larger sample of BBBFL.  They
also found significant, if weaker, correlations between 1500--V and
central velocity dispersion or luminosity.  A later version of the
\mgii\ correlation, including data on Galactic globular clusters, is
shown in Figure 6.  The figure emphasizes the lack of continuity
between the clusters and galaxies (see \S 6.1).  Clusters with a wide
range of Mg$_2$ index have large, if scattered, FUV fluxes.  Some
strong-lined clusters, e.g. 47 Tuc, are faint in the FUV while others,
e.g. NGC 6388 and 6441, are bright.  The galaxies with line strengths
comparable to the strong-lined clusters are relatively faint in the
FUV, but galaxy 1500--V colors rapidly become bluer as Mg$_2$
increases.  The apparent correlation between 1500--V and Mg$_2$ is
much stronger for the galaxies than the clusters.  FUV behavior is
only one of a number of basic spectrophotometric distinctions that
show that globulars and E galaxies do not form a simple population
continuum (e.g.\ Burstein et al 1984, Rose 1985, Ponder et al 1998). 

Using the recent compilation of data for the Lick Observatory E galaxy
spectral survey by Trager et al (1998), one can explore correlations
between the UVX and other absorption line indices.  There are good
correlations, similar to that in Figure 6, between 1500--V and the Na I
D lines or the CN bands at 4150 \AA.  But there is no correlation with
a composite Fe index based on features at 5270 and 5335 \AA.  It is
now clear that the abundances of certain light elements (N, Mg, Na)
are decoupled from those of the iron peak in more luminous E galaxies
(see reviews by McWilliam 1997 and Worthey 1998), although there is
no clear understanding of the nucleosynthetic origin of these
abundance ratio variations.  The behavior of the UVX is evidently
linked to that of the lighter elements.  

The scatter in 1500--V at a given Mg$_2$ is appreciable, especially
among the most metal-rich galaxies.  This may indicate the influence
of parameters other than metal abundance (see \S 6.3).  It is also
possible that the apparent correlation between UV colors and
\mgii\ may not reflect smoothly varying properties but instead might
arise from several discrete classes of galaxies, as discussed by DOR.
There is a suggestion of grouping in Figure 6 (though this is less
pronounced in the correlations with Na I and CN).  Most of the galaxies
have colors in the range $3.5 \pm 0.5$; within this group there is
only a mild UV-Mg correlation.  A few objects, including M32, have
significantly redder colors.  At the other extreme, there are four
strong-lined objects with 1500--V$ < 3$ which stand out as a distinct
group.  

Interestingly, Longo et al (1989) have pointed out that the strongest
UV upturns occur in objects with ``boxy'' isophotes.  Most of the
systems in the middle group of Figure 6 have ``disky'' isophotes.  The
isophotal distinctions between the two groups are now known to
correlate with a wider set of morphological and kinematic properties
(e.g.\ Jaffe et al 1994, Faber et al 1997).  The boxy galaxies are
probably merger products (Bender 1988).  It is therefore possible
that the UVX is influenced by the dynamical environment of galaxies.
Alternatively, all of these characteristics may be related independently
to the mass of galaxies.

In low resolution photometry of 40 early-type galaxies in the Virgo
cluster, Smith and Cornett (1982) detected the effects of the
long-wavelength tail of the UVX component in the integrated mid-UV
colors (2400--V) of E galaxies.  These, however, had a significantly
different color-luminosity relation than the S0 galaxies in the
sample.  Such potential morphological dependencies have not been
carefully investigated.

The UV-\mgii -Na-N correlation and the large internal UV color
gradients (\S 4.2) remain the most suggestive clues linking the UVX to
the global properties of galaxies.  Interpretations of the UVX must
accommodate such correlations, but with the caveat that we do not yet
really understand the nucleosynthetic drivers of E galaxy chemistry.

\section{CANDIDATE LOW MASS UVX SOURCES}

On the basis of the evidence in the last two sections, old, low-mass
stars are the primary sources of far-UV light in E galaxies.
Interpretational effort during the last 10 years has therefore
focussed on the viability of various types of hot, low-mass stars and
their relationship to the dominant populations of E galaxies.  Since
observational information on the UVX is still sparse, much of this
work has been based on new generations of theoretical models for
advanced stellar evolution.  A fully satisfactory interpretation has
not emerged, but there has been good progress in narrowing the range
of possibilities.  In this section and the next we review the main
conclusions.

\subsection{\it Globular Cluster-Type Populations}

The presence of hot stars in old populations was, of course, not
unprecedented because ``blue horizontal branch'' (BHB) stars had long
been associated with metal-poor globular clusters (having $ {\rm
[Fe/H]} \la -1$).  The most natural old-star interpretation of the UVX
was therefore that it arose from the low-metallicity tail of a stellar
population with a large abundance range.

Surprisingly, however, observations quickly demonstrated that the UVX
could not simply be the sum of globular cluster-type populations.  The
first quantitative comparison between clusters and a UVX source was
made using ANS data for the bulge of M31 by Wu et al (1980).  In
order to fit the far-UV spectrum of M31, Wu et al were forced to add
an additional high temperature component to the cluster M13, which has one of
the hardest UV spectra among Galactic globulars.  Models by Nesci \&
Perola (1985) likewise showed that normal cluster BHBs could not
match the galaxy IUE spectra unless a second, hotter HB component was
included.  Oke et al (1981), Welch (1982), and Bohlin et al (1985)
all emphasized the dissimilarity between typical cluster and galaxy
UV energy distributions.  Compilations of ANS, IUE, HUT, and UIT data
for globulars (van Albada et al.~1981, Castellani \& Cassatella 1987,
Bica \& Alloin 1988, Davidsen \& Ferguson 1992, Dorman et al 1995) show
that although clusters in general have total UV-bright star fractions 
exceeding those of galaxies (see Figure 6), galaxies can have steeper
far-UV spectra than any cluster.  The distinction between the two
populations is quite clear in two-color diagrams such as Figure 7,
where the color 1500--2500 is used to measure the slope of the UV
upturn.

\begin{figure}
\plotfiddle{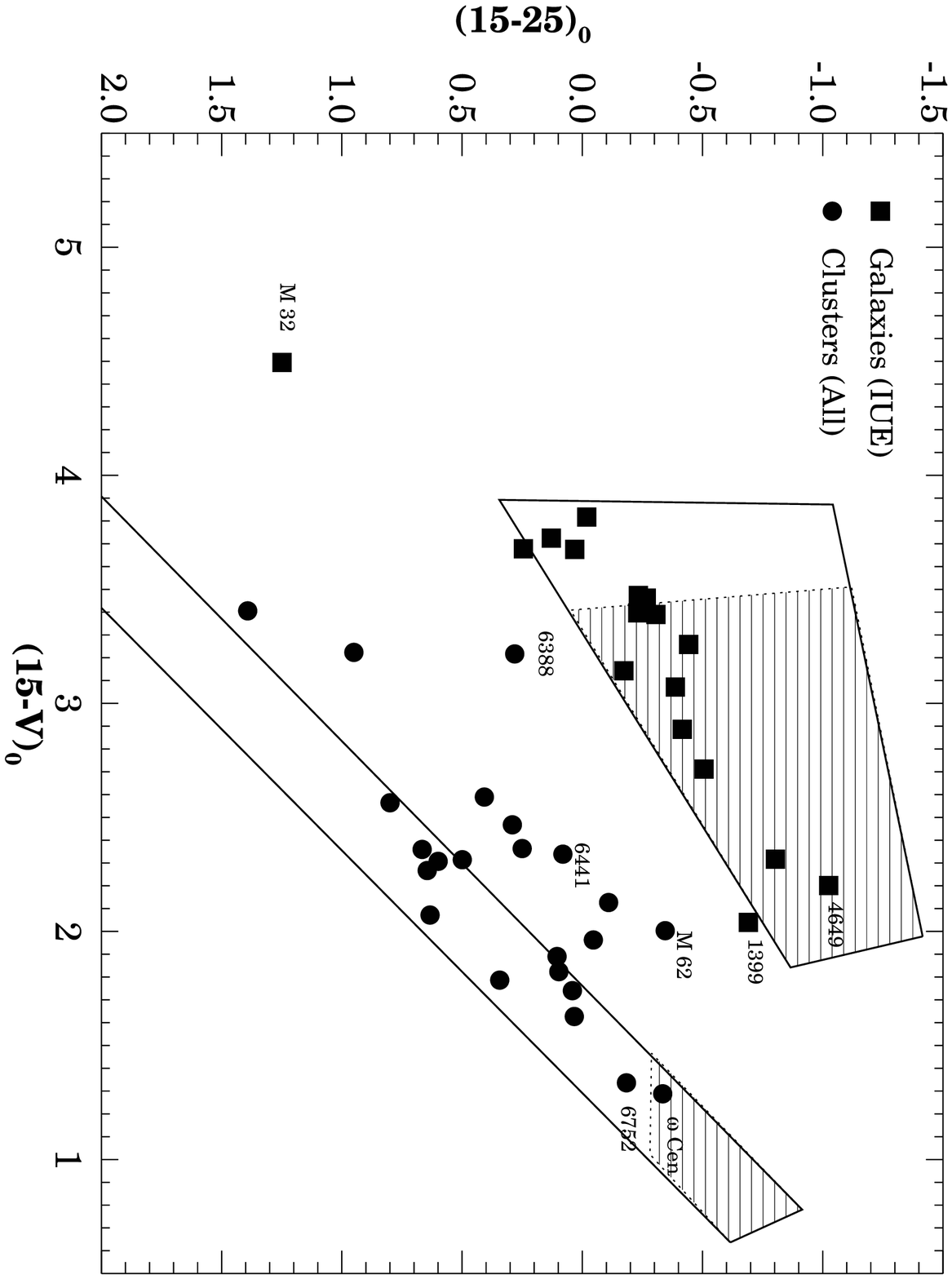}{4.0truein}{90}{60}{60}{+240}{-38}
\caption{
 Broad-band UV colors for E galaxies and
 globular clusters compared.  Data are mainly from IUE.  The color
 1500--2500 measures the slope of the UV-upturn component, while
 1500--V measures its amplitude.  Globular clusters and galaxies are
 clearly segregated in the diagram, with galaxies having steeper FUV
 spectral slopes at a given amplitude.  The boxes enclose several
 fiducial model sets, with the upper one corresponding to $Z \ga \zsun$
 and the lower to $Z \la 0.04 \zsun$.  The shaded regions represent
 models in which the hot stellar component consists mainly of EHB
 stars.  Most of the galaxies require an EHB contribution, but most of
 the clusters do not.  From Dorman et al (1995).
}
\end{figure}

The temperature distribution of UV-bright sources in the clusters is
evidently cooler and broader than in the galaxies.  The far-UV slope
in the galaxies is equivalent to $\teff \ga 20000\,$K, whereas BHBs
in Galactic globulars normally do not extend beyond $\teff \sim\,$
10000--12000 K.  Hotter stars can be found in some globulars on or
above the horizontal branch, as Hills (1971) originally pointed out,
but these are relatively uncommon (see de Boer 1985, 1987 and
\S 6.2).  Even where hotter stars are present, the mean integrated
UV light of typical clusters is heavily influenced by cooler
horizontal branch objects with a wide range of temperatures.  Lower
metallic line blanketing in the atmospheres of cluster stars on the
warm HB and near the main sequence turnoff also produces larger mid-
and near-UV emergent flux than for galaxies, tending to flatten the UV
energy distributions.

Even if one ignores the distinctions between clusters and galaxies in
UV spectral shape, the limits on metal-poor light at optical
wavelengths in M31 and E galaxies are inconsistent with a large
contribution from cluster-like populations in the UV (O'Connell 1976,
1980; Rose 1985; Bica \& Alloin 1988; Dorman et al 1995).

The fact that the galaxy UVX does not appear to arise from a
metal-poor subpopulation similar to the Galactic globular clusters does
not, of course, necessarily mean that the kinds of hot stars in the
galaxies differ from those in the clusters---only that the mixture of
these is different.

\subsection{\it Single Star Candidates}

Since there were no ready-made local analogues for UVX
populations, it was necessary to explore alternatives from a largely
theoretical perspective.  The seminal discussion of the various
low-mass candidates for the UVX sources was presented by Greggio \&
Renzini (1990, hereafter GR; an updated overview is in Greggio
\& Renzini 1999).   They discussed primarily single-star
candidates, since the parameter space for binaries is much larger.
Here, we also defer discussion of the possible involvement of binaries
until a later section.

On the basis of the UV/optical colors, the UVX is estimated to
contribute $\sim$2--3\% of the bolometric luminosity in the most metal
rich galaxies such as NGC 1399 and 4649 (GR).  While this is not
large, the challenge is to identify a mechanism for producing
sufficient numbers of high-temperature stars in old populations.  If
the relevant evolutionary phase is short-lived compared with the
lifetime of the galaxy, then the number of stars in this phase is
proportional to its evolutionary lifetime.  In this circumstance, the
``fuel consumption theorem'' shows that its contribution to the
bolometric luminosity of the galaxy is proportional to the total
amount of nuclear fuel consumed during the phase, which can be
estimated directly from interiors models (Tinsley 1980, Renzini 1981a,
Renzini \& Buzzoni 1986, GR).  Interesting candidates for the UVX
will therefore have temperatures over $\sim 20000\,$K and will burn up
to $\sim 0.01\,\msun$ of hydrogen or $\sim 0.1\,\msun$ of helium
(GR).  Dorman et al (1995, hereafter DOR) used a similar approach to
estimate that the integrated monochromatic energy release at 1500 \AA\
of suitable candidate evolutionary phases must be $\efifteen \ga
4\times 10^{-3}$
\Lvsun\ Gyr ${\rm \AA}^{-1}$, where $\Lvsun = 4.51\times 10^{32}$
\ergsec. 


\begin{figure}
\plotfiddle{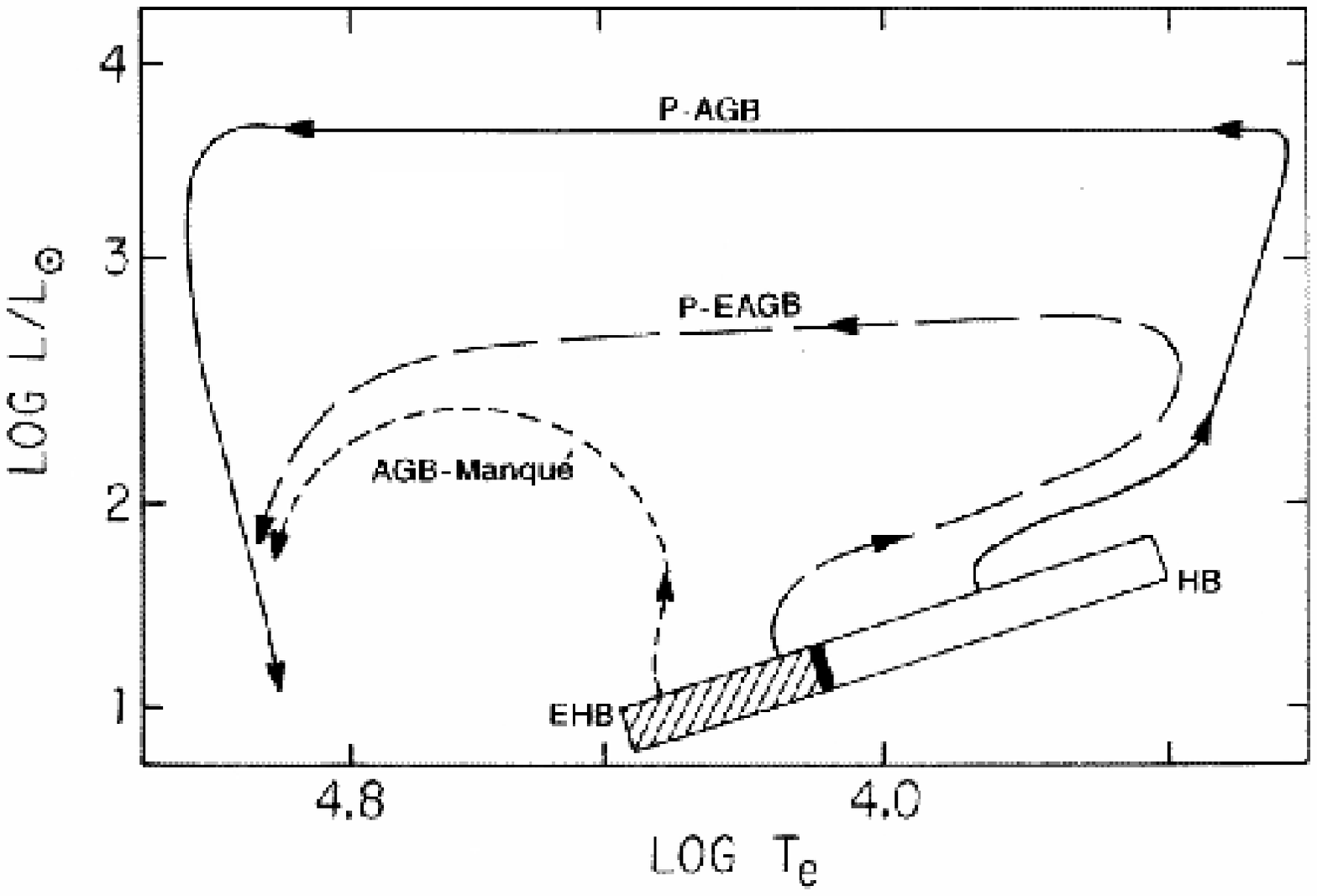}{3.5truein}{0}{80}{80}{-240}{-185}
\caption{
Schematic evolutionary tracks for the
principal post-horizontal branch evolutionary phases described in \S
6.  Envelope masses on the horizontal branch increase from left to
right.  For $Z \sim \zsun$, they are $\menv \sim 0.003 \msun$ at the
left hand (hot) edge and $\sim 0.05 \msun$ at the heavy separator
line, which marks the cool end of the ``extreme horizontal branch''
(shaded).  The segment of the ``P-AGB'' track (solid line) which rises
to the right of $\log \teff \sim 3.6$ corresponds to the AGB. Detailed
evolutionary tracks for these phases are shown, for example, by Dorman
et al (1993).
}
\end{figure}

The evolutionary phases of interest all occur after a low-mass star has
begun moving up the red giant branch.  HR diagram loci for the main
types of UVX candidates that have been explored to date are
illustrated in Figure 8.  The general considerations are described in
GR.  The evolutionary trajectory of a low-mass star following He core
ignition at the tip of the red giant branch is governed mainly by its
envelope mass, $M_{ENV}$.  Since the He core mass is $\sim 0.5 \msun$
and is relatively insensitive to other parameters, $M_{ENV} \sim
M_{TO} - \Delta M - 0.5
\msun$, where $M_{TO}$ is the turnoff mass and $\Delta M$ is the total
mass loss during the red giant phase, which, in globular cluster
stars, amounts to $\sim$ 0.1--0.3 \msun, or 10--40\% of the initial
mass.  The variance in mass loss leads to a scatter in
$M_{ENV}$ and hence in the initial temperature of the subsequent core
He-burning stage on the ``zero-age'' horizontal branch (ZAHB).
Envelope masses on the HB range up to $\sim 0.4 \msun$.  The lower is
$M_{ENV}$, the hotter is the ZAHB location (e.g.\ Iben \& Rood 1970,
Rood 1973, Sweigart 1987, Dorman 1992).  Envelopes smaller than 0.05
\msun\ correspond to \teff({ZAHB}) $\ga 14000\,$K.  After $\sim 100$
Myr, helium becomes exhausted in the center of the star, which then
contains both helium-burning and hydrogen-burning shells moving
outward.

If $M_{ENV}$ is large enough, post-HB stars develop a deep convective
envelope, evolve to lower temperatures, and ascend the cool asymptotic
giant branch (AGB), leaving it only at high luminosity near the AGB
tip, when rapid mass loss and thermal shell pulsing remove the
envelope.  Subsequent evolution in this case involves a rapid
(10$^4$--10$^5$ yr) contraction and heating (the post-AGB or PAGB
phase), in some cases with the formation of a planetary nebula,
followed by cooling and fading on the white dwarf remnant sequence.
Much of the pre-white dwarf time is spent at high temperatures, $\teff
> 50000\,$K.  PAGB models for low-mass stars have been computed by
Sch\"onberner (1983), Bl\"ocker \& Sch\"onberner (1990), and
Vassiliadis \& Wood (1994).  The great majority of stars now near or
above the main sequence turnoff in globular clusters will pass through
the PAGB channel.

More exotic evolution can occur in the case of very small envelopes.
For $M_{ENV} \la\, 0.05\, \msun$ the post-HB star may evolve to higher
temperatures before it reaches the AGB tip or even before it
approaches the cool AGB.  These cases produce, respectively,
post-early AGB (PEAGB) and \manq\ (``failed AGB'') stars (see
Figure 8).  The first detailed models were described by Brocato et al
(1990) and Caloi (1989), respectively (though similar behavior had
been noted in early models by Sweigart et al 1974 and Gingold 1976).
Their internal structure is similar to that of an AGB star (a double
shell source) but with much thinner envelopes.  They burn about the
same amount of fuel as do the more familiar cool AGB stars, but they
do so at much higher \teff\ ($\sim 25000\,$K).  Their post-HB paths in
the HR diagram can be convoluted.  AGB-manqu\'e lifetimes are $\sim
10^7$ yr, considerably longer than for the PAGB phase, after which stars
evolve directly to the remnant cooling sequence.  Typical
\efifteen's for these ``slow blue'' post-HB phases (Horch et al 1992)
are comparable to those of the hot HB phase (DOR Figure 6).  Grids of
such models, for a wide range of metallicities, have been computed by
Castellani \& Tornamb\`e (1991), Castellani et al (1992), Horch et al (1992),
Dorman et al (1993), Bertelli et al (1994), Castellani et al (1994),
and Yi et al (1997a).

The least massive envelope ($\sim 0.05\, \msun$, if $Y \sim
Y_{\odot}$; see Dorman et al 1993, Table 1) capable of producing a
classical PAGB star yields a boundary on the ZAHB between what is now
called the ``extreme HB'' (EHB) (to higher temperatures) and the
normal HB.  This is marked on Figure 8.  \manq\ progenitors occupy
the hot end of the EHB, while PEAGB progenitors 
occupy the end adjacent to the
normal HB.  Note that the normal main sequence for massive stars (not
shown) crosses the HB locus at $\teff
\sim 10000\,$K, and that HB objects hotter than this fall below
the main sequence in the classical ``hot subdwarf'' regime.

Another variety of hot star can be produced directly from the first
ascent red giant branch if mass loss is large enough to remove the
convective envelope before core He ignition.  In this case, the
post-RGB (PRGB) object evolves rapidly to the white-dwarf cooling
sequence without passing through the HB phase.  Some such objects may
experience a late He-flash while on the cooling sequence as their
central temperatures rise owing to gravitational core contraction.
These ``hot flashers'' will then move to a position slightly below the
EHB and follow subsequent post-EHB  tracks similar to
normal EHB stars.  The hot flash effect was demonstrated by Castellani
\& Castellani (1993), and more detailed models including the secular
effects of mass loss during advanced RGB evolution have been computed
by D'Cruz et al (1996).

Remnants on the white dwarf cooling sequence are the inevitable
descendents of all the preceding types of stars.  During their early
evolution, white dwarfs are still hot enough to emit UV photons, and
their potential contribution to galaxy light can be estimated by
integrating down the cooling curve.  Magris \& Bruzual (1993) and
Landsman et al (1998) find that hot white dwarfs (residing on the
cooling curve for $\la 200$ Myr) can contribute up to $\sim 10\%$ of
the far-UV light produced by their parent PAGB phases (less if their
parents were EHB stars).  This is too small to be of practical
importance in normal circumstances.  Unless one invokes an IMF
truncated below $\sim 1.5\,\msun$, the integrated spectrum of the
cooling curve also does not contain the strong Ly-$\alpha$ satellite
features claimed by Bica et al (1996) to be present in IUE spectra of
galaxies.

Finally, the much-studied ``blue stragglers'' (Bailyn 1995), which
are warm stars lying near the main sequence but above the turnoff
luminosity in star clusters, are too cool to be viable UVX
candidates.  They generally have temperatures below 10000 K.  However,
they may influence the mid-UV spectrum of old populations (Spinrad et
al 1997, Landsman et al 1998).  

A common characteristic of the viable UVX candidates is their extreme
sensitivity to small changes in properties.  Differences of only a few
0.01 \msun\ in envelope mass for hot HB stars produce large changes in
the type of post-HB track followed and the resulting \efifteen\ (e.g.
see DOR Figure 6).  Likewise, models for PAGB stars show that their UV
output is extraordinarily sensitive to core mass.  Sch\"onberner's
(1983) 0.546 \msun\ model has a lifetime 20$\times$ longer than for
his 0.565 \msun\ model and has \efifteen\ a factor of 6.8 larger (DOR;
GR Figure 3).  

Likely individual examples of all these candidate types have been
found in local star clusters and the Galactic field.  Imaging with
space telescopes has produced a fairly large sample of PAGB, EHB,
\manq, and related post-HB stars in some globular clusters (e.g.\ in
$\omega$ Cen, Whitney et al 1994, Whitney et al 1998; NGC 6752,
Landsman et al 1996; NGC 2808, Sosin et al 1997; NGC 6338 and 6441,
Rich et al 1997; M13 and M80, Ferraro et al 1998).  A smaller sample
of similar sources has been identified in the open clusters NGC 188
and 6791 (Liebert et al 1994, Landsman et al 1998), and Landsman et al
estimate that NGC 188 and 6791 would have UV upturns in their
integrated light as strong as any E galaxy.  Over 1500 hot subdwarfs
(sdO, sdB, and related types) are now known in the Galactic field.  As
first shown by Greenstein \& Sargent (1974), many of these are EHB and
post-EHB stars (Heber 1992, Saffer et al 1994).  Kinematical studies
show that some hot subdwarfs are members of the old, metal-rich disk
population of the Galaxy (Thejll et al 1997).  The field and open
cluster examples are important cases since they demonstrate that EHB
objects are not confined to low-metallicity environments.  NGC 6791,
in particular, has [Fe/H] $\sim +0.5$ (Kaluzny \& Rucinski 1995).

DOR summarized integrated UV outputs for the several main candidate
UVX star types.  PAGB tracks have $\efifteen < 0.001$ \Lvsun\ Gyr
${\rm \AA}^{-1}$.  They therefore cannot be solely responsible for the
brightest UV-upturn cases, as first pointed out by GR and Castellani \&
Tornamb\`e (1991).  However, the EHB, PEAGB, and \manq\ phases
burn more H$\,+\,$He fuel at high $\teff$'s, by factors of $\sim$
3--30, than classical PAGB stars and therefore are excellent UVX
candidates.

\subsection{\it Relationship to Global Characteristics of Parent Population}

Each UVX candidate represents a potential channel to be filled by its
evolving post-main sequence parent population.  At least five global
population parameters are known to be important in determining the
occupation of the various channels.  The effects of these have been
reviewed in GR and Chiosi (1996).  Yi et al (1997b) nicely illustrate
the effects of age, abundance, and mass-loss parameters on
color-magnitude diagrams, integrated spectra, and broad-band UV
colors.  The single most important variable is mass loss on the giant
branch, followed by helium abundance ($Y$).

\noindent{\it Age:} As a population ages, its turnoff mass decreases,
with $M_{TO} \sim 0.96\, {t_{10}}^{-0.2}\, \msun$, where $t_{10}$ is the
age in units of 10 Gyr.  For a given amount of RGB mass loss, older
stars will have smaller $M_{ENV}$ and will fall at higher
ZAHB temperatures.  The UVX is therefore expected to increase with
age, though probably in a strongly nonlinear fashion.  At large enough
ages, all stars evolving up the RGB will become hot EHB or PRGB
objects.  

\noindent{\it Y:} An increase in helium abundance has important effects on
post-giant branch evolution (GR, Horch et al 1992, DOR).  
Because of the increase in mean molecular weight, turnoff masses at a
given age are smaller, which yields smaller $M_{ENV}$ for a given
amount of RGB mass loss.  A higher initial helium abundance also 
causes stars with a given $M_{ENV}$ to burn more of their hydrogen
envelope during the core He-burning phase, producing \manq\
behavior for a larger range of $M_{ENV}$'s (Horch et al Table 1; DOR
Figure 6).   Increasing Y
from 0.27 to 0.47 roughly quadruples the total \efifteen\ for a
uniform distribution of $M_{ENV}$'s (DOR).  

\noindent{\it Z:} The strong correlation between the UVX and line
strengths discussed in \S 5.2 makes metal abundance effects on hot star
evolution of particular interest.  Based on the example of the
globular clusters, one might suppose that metal abundance determines the
prevalence of hot HB stars.  To the contrary, theoretical models show
that Z has little direct effect on either the EHB or post-EHB phases
of evolution (e.g.\ Dorman et al 1993, DOR).  Instead, these are
governed mainly by $M_{ENV}$.  Although increased metallicity does
increase $M_{TO}$ for a given age, thereby decreasing ZAHB
temperatures for a given amount of RGB mass loss, hot HB stars can appear
at any metal abundance as long as envelope masses are small enough.  This is
demonstrated in the grids of metal-rich HB models cited in \S 6.2
and was first illustrated in integrated light  
by Ciardullo \& Demarque (1978).
However, for higher metallicities, \teff\ for medium-envelope
(0.05--0.15 \msun) stars is strongly decreased.  This implies that a
uniform distribution of $M_{ENV}$ will lead to a bimodal distribution
of ZAHB temperatures at higher metallicities (Dorman et al 1993,
D'Cruz et al 1996). 

There has been less exploration of advanced evolution with relative
abundance variations among the metals.  D'Cruz et al (1996) found no
qualitative changes in behavior for models with ${\rm [O/Fe]} =
+0.75$.  However, as discussed in \S 5.2, models incorporating
variable abundance ratios among the metals would seem to be essential
if the empirical line strength correlations are to be understood.  

These theoretical expectations on the secondary status of metallicity
effects on HB temperatures have good empirical support.  The bluest UV
colors in Figure 6 occur not for the most metal-poor globular clusters
but for those of intermediate metallicity.  Small numbers of EHB and
related stars have recently been found in globular clusters with
heavily populated red HBs (NGC 362, Dorman et al 1997; 47
Tucan\ae, O'Connell et al 1997), and large numbers of hot HB stars are
present in the relatively metal-rich clusters NGC 6388 and 6441 (with 
$Z \sim 0.25 \zsun$, Rich et al 1997).  Other clusters with
EHB stars may range up to $Z \sim 3\zsun$ (NGC 6791,
Liebert et al 1994).

\noindent{ $\Delta Y/\Delta Z$}:  There is good evidence from the
study of emission lines in low-metallicity galaxies that helium
abundance is coupled to metal abundance (e.g.\ Wilson \& Rood 1994,
Izotov \& Thuan 1998).  Values of $\Delta Y/\Delta Z \sim 3$--4 have
been derived for low metallicity environments.  If these apply to E
galaxies, then the smaller effects on EHB and post-HB evolution of
metallicity enhancements for $Z \ga Z_{\odot}$ are strongly amplified
by the effects of $Y$ enhancement, as emphasized by GR.  The dramatic
increases in post-HB UV output found by Horch et al (1992) in
metal-rich models were actually produced by the accompanying He
effects ($Y$ is increased to $\sim 0.35$--0.45 for $Z > 2\,Z_{\odot}$ in
their models).  J\o rgensen \& Thejll (1993) estimated that $\Delta
Y/\Delta Z > 2.5$ is needed to produce a strong positive correlation
between metal abundance and UVX above
\zsun, for normal ranges of age and RGB mass loss.  DOR (\S 8.3)
emphasize that there is very little known about $\Delta Y/\Delta Z$
for solar abundances or above and that most available chemical
evolution models suggest smaller He enhancements than for low
abundances.  DOR also point out that EHB stars exist in clusters and
the Galactic field at moderate metallicites, and presumably moderate
$Y$s, so that extreme Z or $Y$ enhancements are not essential to their
production.

\noindent{\it Mass Loss}:  Mass loss is the most important
determinant of post-RGB evolution in low-mass stars but is also
the most difficult to evaluate because of a paucity of both
empirical evidence and theoretical exploration.  RGB mass loss
is usually modeled using the Reimers (1977) prescription:

$$ \dot{m} = -4 \times 10^{-13}\: \etar\:\frac {L}{g R} \:\: M_\odot\,
{\rm yr}^{-1}$$

\noindent where $\etar$ is a mass loss efficiency parameter, $L$ is
the luminosity, $g$ is the surface gravity, and $R$ is the radius,
with $L,~g,~ {\rm and}~R$ in solar units.  This formula is based on
dimensional analysis rather than a well-grounded physical theory.  It
is consistent with the available observational data, which indicate
only that mass loss increases with luminosity and decreasing surface
temperature, reaching a maximum just prior to the He flash (reviewed
in Dupree 1986).  The Reimers prescription implicitly includes
composition and age dependences through their influence on stellar
structure, and hence $L,~g,~ {\rm and}~R$.  Although \etar\ is the
principal mass-loss parameter in this formulation, empirically there
is always a significant spread in the effective \etar\ (e.g.\ Rood 1973), which
produces a range $\Delta\menv$ on the ZAHB.  There is no theory for
the spread at the moment, so it appears in evolutionary synthesis
models as an additional free parameter.

To produce a typical globular cluster blue horizontal branch requires
$\etar \sim 0.2$--0.5 (e.g.\ Renzini 1981b, Yi et al 1997b) if cluster ages
are $\sim 15$ Gyr.  Such values are therefore regarded as ``normal''
and are widely used in galaxy spectral modeling.  However, the
globular cluster values would increase if the lower ages
of $\sim 12$ Gyr favored by the recent Hipparchos recalibration of
distance indicators are adopted (Yi et al 1999).  Furthermore, there
is very little evidence on whether globular cluster
\etar\ values are appropriate in other kinds of stellar populations.
It is physically plausible that
\etar\ would increase with metal abundance, owing to grain formation,
for instance.  GR explored the consequences of assuming that $\etar
\propto 1 + \frac {Z}{Z_{crit}}$, where $Z_{crit}$ is a critical
threshold.  D'Cruz et al (1996) and Yi et al (1997b) considered models
for a range of \etar\ up to 1.2, the former including self-consistently
the effects of mass loss on evolution near the RGB tip and the ``hot
flasher'' phenomenon.  They find that production of EHB stars in
populations with $Z \ga \zsun$ requires $\etar \ga 0.7$, with total
mass loss of $\sim 0.5 \msun$ per star for ages $\ga 10$ Gyr.  D'Cruz
et al found that a smooth distribution of \etar\ on the RGB leads to a
strongly bimodal distribution of ZAHB temperatures if $Z \ga \zsun$.  
The models imply that increases of \etar\ of only a factor of 2--3
over canonical globular cluster values are sufficient to produce
a large population of EHB and post-EHB stars for normal ranges
of age, $Z$, and $Y$.  

The hot flash phenomenon is one way of routing a significant
fraction of the evolving population into HB objects with $\menv < 0.05
\msun$ without the need for ``fine tuning'' of mass loss. D'Cruz et al
(1996) showed that as long as mass loss near the RGB He flash
luminosity is above a critical threshold (corresponding to $\etar
\sim 0.7$, but not necessarily tied to the Reimers prescription), EHB
stars will always be produced via the hot-flash mechanism.  Neither
the \etar\ values nor the range of values ($\Delta\etar$) producing
EHB objects vary much with metallicity from globular cluster to
supersolar values.  The mechanism produces a natural concentration
to a narrow range of $\teff$.    There is also improving empirical evidence from
the cluster and field samples (\S 6.2) that EHB
populations do occur in nature at a wide range of metallicities.
Thus, there are no obvious obstacles to the production of EHB
stars through enhanced mass loss.  

Although the Reimers law provides a useful schematic for exploring
mass-loss effects on the UVX, a much improved physical theory is
needed.  Preliminary hydrodynamic models (e.g.\ Bowen \&
Willson 1991, Willson et al 1996) suggest a sudden onset of mass loss
at a critical luminosity and a strong metallicity dependence, effects
that may not be well modeled by a simple scaling law.

\noindent{\it Other Parameters:} There are certainly other processes
that can influence the production of hot stars in old populations.
Sweigart (1997) showed that deep mixing in the outer envelopes of RGB
stars, which results in enhanced surface He abundances, encourages the
production of hot HB stars and \manq\ behavior.  Larger mixing is
presumably related to higher stellar rotation rates.  The dynamical
environment of galaxies could therefore influence the UVX by way of
stellar spin distributions.  Ferraro et al (1998) find that gaps in
the hot HBs of different globular clusters occur at similar
temperatures, suggesting that RGB mass loss is a multimode process.
Good candidate mechanisms have not yet been identified.

\subsection{\it Binaries and Dynamical Effects}

There has been much speculation about the possible origin of hot
low-mass stars through dynamical interactions, especially in binary
systems through Roche-lobe mass transfer or mergers.  The various
recognized mechanisms have been reviewed by Bailyn (1995), while their
implications for the UVX problem were most extensively discussed by
GR.  The mildest form of interaction occurs when a star ascending the
giant branch loses part of its expanding envelope to a companion,
thereby appearing with lower \menv\ on the ZAHB (Mengel et al 1976) but
evolving normally thereafter.  Based on the frequency of binaries
among the sdB stars in the Galactic field (Green 1999), this
process may be fairly common there. Although ``fine tuning'' of binary
mass ratios and separations would seem to be necessary to produce
small \menv\ without suppressing the He flash altogether, in fact the
hot-flash mechanism (D'Cruz et al 1996) would mitigate this problem
here as it does for normal mass loss.  One reason that UV star
production in binaries might depend on $Z$ is that stellar envelopes
become more inflated at higher metallicities (GR).  The fact that
field and cluster hot horizontal branches have gravities and
luminosities consistent with single-star models suggests that more
drastic interactions (e.g.\ mergers) are considerably more rare.

Some support for dynamical effects is provided by the observation that
the extent of horizontal branch ``blue tails'' in Galactic globular
clusters appears to correlate with cluster concentration and density
(e.g.\ Fusi Pecci et al 1993, Buonanno et al 1997).  However, other
expectations for dynamical mechanisms are not met.  For instance, the
hot stars are not necessarily concentrated to the centers of clusters,
and the system with the largest EHB/post-EHB population ($\omega$ Cen)
is notably low density (e.g.\ Whitney et al 1994, Rich et al 1997).
HST observations of 10 cluster cores (Sosin et al 1997) show that the
EHB stars are not as centrally concentrated as the blue stragglers
(which are almost universally agreed to be interaction products).

It is unclear how to translate the evidence for dynamical mechanisms
in star clusters to the dynamical environment of galaxies, which is
very different and less conducive to stellar interactions.  Of course,
the present field population in E galaxies may well have originated in
concentrated cluster-like systems that have since disintegrated but
which were responsible for establishing the binary frequency.  Stellar
rotation, which could depend on global dynamical characteristics of
galaxies, may also influence the UVX through He mixing (Sweigart
1997).  The only hint that the E galaxy UVX is related to dynamics is
the connection between large UV upturns and boxy isophotes (Longo et al
1989 and \S 5.2 ).  The core of M32 is the densest observable
extragalactic system.  No large radial gradients in blue light are
apparent near its center at HST resolution (King et al 1992, 1995;
Bertola et al 1995, Cole et al 1998, Lauer et al 1998), though Brown
et al (1998a) note that the density of resolved UV stars per unit total
light increases slightly at smaller radii in both M31 and M32.

Mass transfer onto white dwarf binary companions can produce copious
UV flux and has also been discussed in the context of the UVX by GR.
One of the main problems is again the fine-tuning needed to ensure a
sufficient but not excessive transfer.  Because of the large parameter
space involved, estimates are only tentative, but accreting white
dwarfs seem unlikely to be major contributors to the UVX (GR).  The UV
spectrum of a population dominated by such objects is also expected to
show fairly strong emission lines (e.g.\ Wu et al 1992), which are
absent in E galaxy spectra.  

Although dynamical interactions could certainly influence the UVX in
galaxies, there is no strong evidence yet that they do so.

\section{HOT LOW MASS STARS IN ELLIPTICAL GALAXIES}

\subsection{\it Interpretation of UV Spectra and Colors}

Early discussions of low-mass candidates for the UVX in the context of
the observations concentrated mainly on distinguishing them from the
massive star interpretation rather than from one another.  Rose \&
Tinsley (1974) were the first to emphasize that hot PAGB stars were
inevitable products of low-mass evolution and should be present in
sufficient numbers to affect the integrated UV spectrum in old
populations of all metallicities (assuming mass loss is not so extreme
as to suppress the AGB phase altogether).  O'Connell (1976) found
tentative evidence for hot starlight in the 3300--4000 \AA\ region
of 3 gE galaxies, which was plausibly interpreted as from the PAGB.
Following the demonstration that normal HB stars in globular cluster-like
populations were not compatible with the galaxy UVX spectra (see
\S 6.1), PAGB stars became the favored candidates.

Bohlin et al (1985), Renzini \& Buzzoni (1986), and O'Connell et al
(1986) pointed out that the extreme sensitivity of the UV output of
PAGB stars to their core masses might explain the large variation in
UVX strength and the Faber (1983) \mgii\ correlation.  They suggested
that metallicity-enhanced mass loss or age differentials
drove the correlation.  BBBFL, Bertelli et al (1989), and Magris \&
Bruzual (1993) examined the implications of PAGB models
quantitatively.  They found that the brightest UVX sources would
require non-PAGB sources of UV light or PAGB stars with core masses
smaller than those in the grid of evolutionary models by Sch\"onberner
(1983), the smallest of which was technically a PEAGB rather than a
PAGB star (i.e.\ it did not reach the AGB tip).  They also
concluded that changes in $Z$ alone could not reproduce the Faber
correlation unless there were accompanying changes in $t$, \etar, or
$Y$.  

Shortly thereafter, two lines of evidence combined to reject PAGB
stars as the principal UVX sources.  First, GR and the other studies
cited in \S 6.2 emphasized the shortfall in the UV output of PAGB
objects compared with the strongest upturn galaxies and suggested EHB
stars and their descendents as a more viable alternative.  Second, HUT
spectroscopy showed that the energy distributions of M31 and NGC 1399
declined shortward of 1050 \AA\ (Ferguson et al 1991, Ferguson \&
Davidsen 1993).  This placed an upper limit of $\teff
\sim 25000\,$K on the temperature of the dominant UVX stars,
considerably cooler than the 50000 K characteristic of a PAGB
component but entirely consistent with the EHB and post-EHB channels.
Ferguson \& Davidsen (1993) found significant differences between M31
and NGC 1399 which demonstrated, independent of modeling details, that
the UVX is probably a composite population with the mixture of HB and
other hot types varying from system to system.  (Later HUT observations
by Brown et al 1997 confirmed such variations in six other galaxies
based on 912--1000 \AA\ fluxes.)

EHB, PEAGB, and \manq\ stars have consequently become the favored
candidates for the dominant UVX sources.  They are energetically
viable since in the brightest UVX cases only 10--20\% of the evolving
stars in the dominant population would need to pass through the EHB
and post-EHB channels to produce the observed far-UV luminosities (GR,
DOR, Brown et al 1995).  DOR showed that the observed 1500--V and
2500--V colors of E galaxies were consistent with composite
EHB/post-EHB and PAGB models in which the EHB channel contributes
$\sim 25\%$ of the FUV light in medium-upturn systems like M31 but
$\sim 75\%$ in the strongest upturns.  They found that most globular
clusters do not require EHB stars to explain their UV colors,
consistent with independent information on color-magnitude diagrams
(see Figure 7).  They also found that if mass loss is left as a free
parameter, 1500--V does not place useful limits on galaxy age or metal
abundance, though 2500--V does.  This is because the properties of the
EHB/post-EHB channels are not very sensitive directly to either
parameter, whereas the main sequence turnoff (which dominates for $\lambda > 2500$
\AA) is.  DOR found acceptable agreement with observed colors for a
wide range of ages (6--20 Gyr) but only for solar or higher metallicities
($Z \sim 1$--4 \zsun), again leaving mass loss unconstrained.

\begin{figure}
\plotfiddle{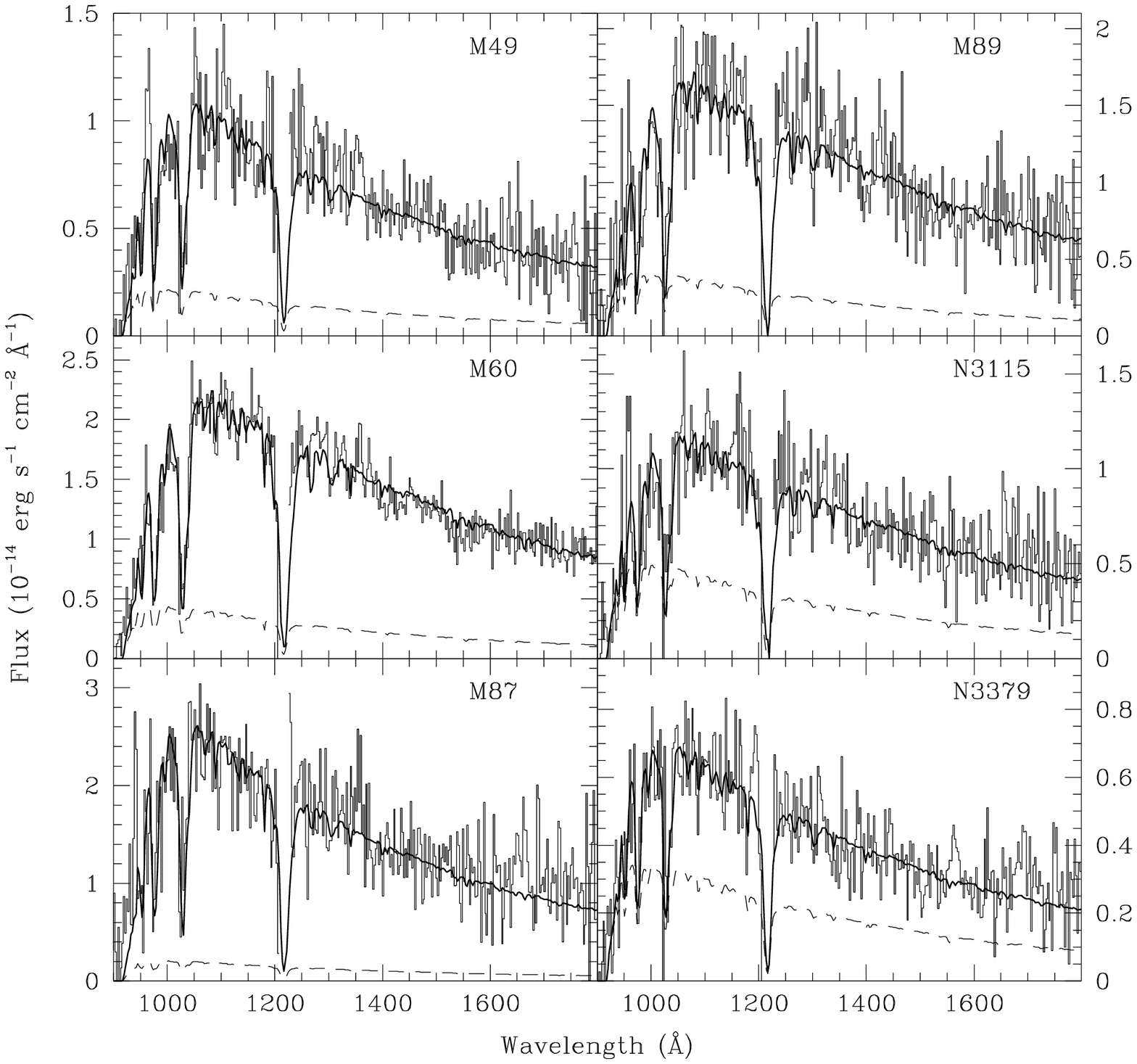}{4.5truein}{0}{65}{65}{-195}{-115}
\caption{
{\it Astro}/HUT far-UV spectra of 6 E/S0
 galaxies compared with EHB$+$post-EHB$+$PAGB models. Fluxes are shown in
 2.5 \AA\ bins.  The best-fitting composite models are shown by the
 solid lines.  These employ evolutionary tracks with $Z
 \sim 2-3 \zsun$ but atmospheres with $Z = 0.1 \zsun$.  The PAGB
 contribution to each model is shown by the dashed lines; this is
 usually considerably smaller than 50\%. From Brown et al (1997).
}
\end{figure}

Brown et al (1997) analyzed HUT spectra of six E/S0 galaxies
covering the 900-1800 \AA\ region at 3 \AA\ resolution.
They were able to produce good fits to the spectra with composite
EHB/post-EHB and PAGB models in which only a small fraction ($\la 10\%$)
of the evolving population need pass through the EHB channel (see
Figure 9).  Although the models formally contained very small ranges
of \menv\ on the ZAHB (e.g.\ 0.021--0.046 \msun\ in NGC 4649), they
found that broader distributions of
\menv\ would yield similar fits because the short wavelength FUV flux
tends to be dominated by the hot \manq\ phases.  The best fits occur
for evolutionary tracks with supersolar values of $Z$ (2-3
\zsun) and $Y$ (0.34-0.45), which produce more flux than subsolar
models below 1200\AA. Interestingly, however, fits are better for
atmospheres with subsolar values of $Z$ ($\sim 0.1 \zsun$).  Brown and
colleagues attribute this to the same processes (mainly diffusion)
that create well-known abundance anomalies among Galactic hot
subdwarfs (e.g.\ Saffer \& Liebert 1995).  These effects are not
straightforward to analyze, but the expectation is that diffusion in
high-abundance stars will tend to reduce line strengths and make
spectra appear less metal-rich.  This is a critical issue, however,
since the consensus interpretation would have to be fundamentally
revised if the UVX arose from a metal-poor population.

The DOR and Brown et al (1997) analyses were consistent with realistic
global population models but left mass loss on the RGB as a free
parameter because the physics involved is so uncertain.  A large body
of other evolutionary spectral synthesis studies adopt definite
prescriptions for mass loss and have predicted the UV spectra of old
populations with the intent of exploring how higher metal abundances
might produce larger UV output (as in Figure 6).  Some of these
consider a fixed grid of abundance parameters while others (e.g.\ the
Padova group, Bressan et al 1994, 1996) include self-consistent star
formation and nucleosynthetic enrichment histories to determine the
abundance distribution in galaxy models of various masses.  In most of
these models, RGB mass loss is specified by the Reimers prescription
using a fixed mean \etar\ value in the range $\sim$ 0.3--0.5, as
appropriate for globular clusters (e.g.\ Bertelli et al 1989, Barbaro
\& Olivi 1989, Bruzual \& Charlot 1993, Magris \& Bruzual 1993,
Bressan et al 1994, Lee 1994, Bertola et al 1995, Bressan et al 1996,
Chiosi et al 1997, Park \& Lee 1997).  These model sets do not include
the ``hot flash'' phase.  It is also necessary to specify the spread
of \menv\ on the ZAHB, $\Delta\menv$.  Since the early studies of
globular cluster HBs (e.g.\ Rood 1973) it has been traditional to use
a modified Gaussian distribution, though this is neither unique nor
well justified on astrophysical grounds.

\begin{figure}
\plotfiddle{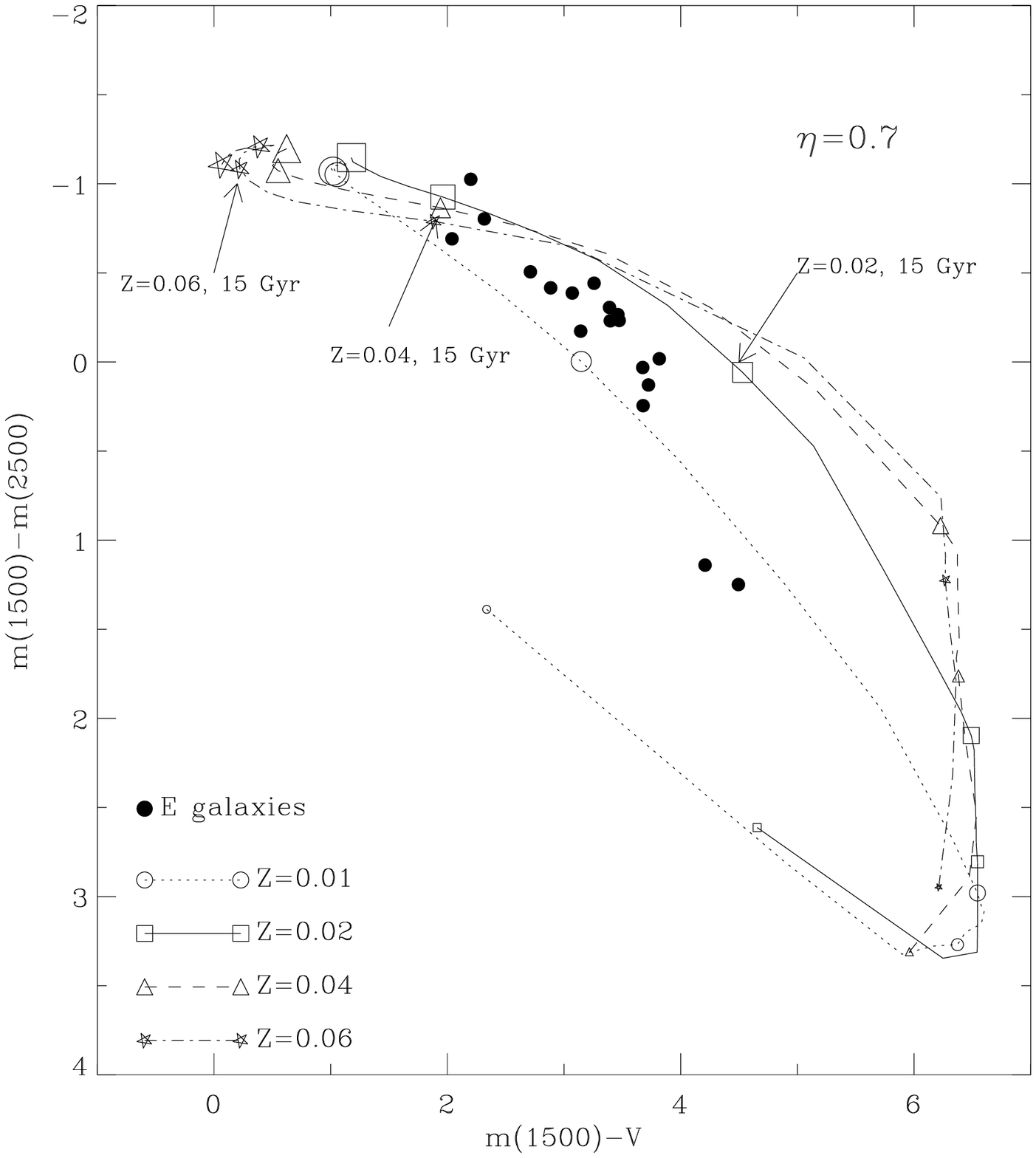}{5.5truein}{0}{80}{80}{-260}{-10}
\caption{
 IUE color data (same axes as Figure 7) for E
 galaxies compared to theoretical models.  From Yi et al (1997b).  The
 various symbol shapes correspond to sequences for different $Z$, as
 indicated on the legend.  The symbol sizes increase with age; the six
 models in each sequence are for 1, 5, 10, 15, 20, and 25 Gyr.  All
 models assume a Reimers mass-loss parameter of $\etar = 0.7$, about
 twice that estimated for globular clusters.  Masses on the ZAHB are
 assumed to be spread about the mean value (determined by \etar) in a
 modified Gaussian distribution with $\sigma = 0.06 \msun$.  $\Delta
 Y/\Delta Z$ is assumed to be 3.0.  See \S 7.1 for further details. 
}
\end{figure}

The two-color diagram in Figure 10 (from Yi et al 1997b) illustrates
the general nature of the predictions from this class of models.  It
shows the effects of age and composition for a grid with $\etar = 0.7$
and $\Delta Y/\Delta Z = 3$, somewhat higher values than typically
assumed in the other studies mentioned above.  The color-age relation
is nonmonotonic.  Far-UV light reaches a minimum at about 5 Gyr, after
the decay of the warm main sequence and before the EHB channel
is well-filled.  PAGB stars are always present but are significant in
the UV light only for ages $\la 5$--10 Gyr, after which the EHB and
post-EHB phases dominate.  The EHB/post-EHB channel becomes strongly
occupied only when the turnoff mass becomes small enough that the
assumed RGB mass loss yields small-envelope ZAHB stars.  In Figure 10
this occurs after $\sim 16$ Gyr for \zsun\ and $\sim 8$ Gyr for
3\zsun, yielding 1500--V $\la 3$.  These threshold ages would decrease
by about 5 Gyr for $\etar = 1$ (Yi et al 1997b, Figure 22).  Although
the detailed age-color relation depends on $Z$, the color loci for $Z
\ga \zsun$ nearly superpose, implying that the UV colors cannot be
easily used to distinguish $Z$.  Tracks for $\etar = 1.0$ reproduce the
plotted E galaxy data better than those shown here, although these do
encompass the entire color range observed.  In particular, the plotted
models allow significantly stronger UV-upturns (1500--V $\sim +0.5$
and 1500--2500 $\sim -1$) than actually observed.  As indicated by the
labeled points, the observed range of 1500--V (though not 1500--2500)
could be produced at $t = 15$ Gyr, $\etar = 0.7$ by a change of $Z$ of
a factor of 2, with a concomitant change in $Y$.  Alternatively, within
the modeling uncertainty, the data could be equally well explained by
changes in age at constant abundance or by a correlation between
\etar\ and abundance.

\subsection{\it Inferences About Chemical Abundances and Ages}

The evolutionary model sets agree that the observed far-UV fluxes can
be produced by old (5--15 Gyr), metal-rich ($Z \sim 1$-3 \zsun)
populations given favorable, but not unreasonable, assumptions about
$\Delta Y/\Delta Z$, \etar, and $\Delta\menv$.  They also agree that
changes in $Z$ alone cannot reproduce the Faber correlation, unless it
is assumed that $\Delta Y/\Delta Z \ga 2.5$ or that \etar\ increases
with $Z$ (as foreseen by GR).  Models based on more conservative
assumptions about mass loss require that populations with significant
UVX components be older than about 10 Gyr.  

As noted in \S 6.3 and illustrated in Figure 10, the metal abundance
$Z$, if decoupled from $Y$ and mass loss, has only a secondary
influence on the spectrum of the hot components at wavelengths above
1200 \AA.  There is a somewhat greater effect on the temperature
distribution of light in the 900--1200 \AA\ range (Brown et al 1997).
$Y$ can have a significant effect on UV flux production, but its
influence can be masked by changes in mass loss.  Furthermore, the UVX
absorption line spectrum is apparently subject to atmospheric
diffusion effects, as discovered by Brown et al (1997).  Therefore
far-UV spectra in the 1200--2000 \AA\ range cannot easily be used to
infer abundances of the hot star populations of E galaxies.

It is also premature to try to use the UVX to age-date elliptical
galaxies.  The far-UV appears promising for this purpose because it is
the most rapidly evolving part of a single-burst galaxy spectrum.  A
number of studies have noted that the ``turn-on'' of the UVX (which
occurs at the age when $M_{TO}$ drops to the point that the assumed
RGB mass loss is able to fill the smaller envelope channels) marks an
obvious spectral transition which might be used to age-date
E galaxies at moderate lookback times (e.g.\ GR, Magris \& Bruzual
1993, Chiosi 1996, Bressan et al 1996, Chiosi et al 1997, Yi et al
1999).  The amplitude of the transition in UV colors is large.
Unfortunately, its timing in the models is very sensitive to
assumptions about mass loss and helium abundance, making results
strongly model-dependent.  This can be seen in the cases presented by
Yi et al (1997b, 1999).

An interesting related example is the interpretation of the UVX as the
product of a metal-poor subpopulation of extremely old (18--20 Gyr)
stars, presumably the earliest generation to form in massive galaxies
(Lee 1994, Park \& Lee 1997).  The models employed by Lee and Park
adopt atypically small values for both total RGB mass loss (a fixed
0.22 \msun) and $\Delta\menv$ (0.02 \msun).  Larger values for these
parameters would significantly reduce the inferred ages.  The models
also do not fit the UVX spectra very well, having the flatter energy
distributions for 1500--3500 \AA\ characteristic of metal poor systems
(Park \& Lee 1997), even when mixed metallicites and larger effective
mass loss are included (Yi et al 1999).

It is clear in general that assumptions about \etar\ and $\Delta\menv$
largely determine the outcome of the far-UV evolutionary synthesis
models developed to date and any conclusions about $t$, $Y$, or $Z$
that may emerge from them.  This is necessarily
so, given the circumstance that changes in only a few 0.01 \msun\ in
\menv\ can radically affect the UV output of stars.  Age and mass loss
can be traded for one another, and without a more deterministic theory
of mass loss, derived ages are not reliable.

These remarks apply to the far-UV, hot-star spectra of old populations.
The situation is quite different in the mid-UV region (2400--3200 \AA)
where the light becomes dominated by cool stars ($\teff \sim
5500$--7500 K) near the main sequence turnoff.  Turnoff light is
directly sensitive to both age and abundance, whereas red giant light,
which is important longward of 4500 \AA, is insensitive to age.  The
principal obstacle to exploitation of the mid-UV as a population
diagnostic is the lack of good ``libraries'' of mid-UV stellar energy
distributions.  Empirical datasets (e.g.\ Fanelli et al 1992) tend to
be limited to solar abundance, whereas theoretical ones (e.g.\ Kurucz
1991) have serious shortcomings due to difficulties in treating the
overwhelming UV line blanketing.  HST is beginning to fill the gap, if
slowly, with high quality, medium-resolution spectra (e.g.\ Heap et al
1998).  Another technical difficulty with the mid-UV is that
contamination by the long-wavelength tail of the UVX energy
distribution must be removed.  Although this contributes over 50\% of
the 2700 \AA\ light in many cases (BBBFL, Ponder et al 1998),
correction for the UVX component is straightforward because it has a
smooth and well-determined shape (Dorman et al 1999).  

Preliminary synthesis models of the mid-UV using theoretical stellar
spectra confirm the expectation that it will be an excellent
age/abundance diagnostic.  DOR's experiments with fitting broad-band
UV colors of E galaxies showed that 2500--V yields much more
information on age and composition than does 1500--V.  They estimate
that $ \partial(2500$--V)$/\partial \log Z \sim 2.7$ for old
populations, a much higher sensitivity than for most optical-IR
indices (e.g.\ Worthey 1994).  The mid-UV continuum is especially
useful in placing limits on the contribution of metal-poor populations
to galaxy light.  The available mid-UV models show, for instance, that
the metal-poor fraction in E galaxies is much smaller than predicted
by simple ``closed box'' nucleosynthetic models (e.g.\ Tantalo et al
1996, Worthey et al 1996).  Empirically, mid-UV spectral features also
strongly distinguish the populations of globular clusters and E galaxy
cores from one another (e.g.\ Ponder et al 1998).  

One of the most important applications of mid-UV stellar population
analysis will be to the spectra of high redshift galaxies.  Age-dating
of distant objects that are passively evolving, such as LBDS 53W091,
with $z = 1.55$, can constrain the earliest epoch of star formation,
and hence cosmology (e.g.\ Spinrad et al 1997).

\subsection{\it Resolved UV Star Populations}

The HST has sufficient sensitivity to probe directly the UV star
populations of nearby galaxies.  Three UV imaging studies of hot stars
in M31 and M32, all based on FOC observations of small fields, have
appeared to date (King et al 1992, Bertola et al 1995, Brown et
al 1998a).  Although the two earlier studies suffered from serious
calibration difficulties (see Brown et al 1998a), all three detected
UV-bright stars and agree that luminous PAGB stars cannot account for
more than a small fraction of the total FUV light.  The photometry of
Brown et al (1998a) has a detection limit of m$_{\lambda}(1750\, {\rm \AA})
\sim 24.5$, which is not deep enough to reach the HB itself but does
encompass PAGB, PEAGB and \manq\ luminosities.  Brown and colleagues
identify a large number of stars consistent with expectations for the
descendents of EHB stars with $\menv \sim 0.002$--0.05 \msun.  However,
most of the UV light is produced by unresolved stars, presumably on
the EHB.  The integrated HUT or IUE spectra are consistent with
models, normalized by the resolved samples, in which only 2\% of the
total population passes through the EHB channel in M31 and only 0.5\%
does so in M32.  The lifetime of the PAGB channel, which makes up the
rest of the post-HB population, is so short that few resolved objects
are expected in the observed fields.  Somewhat unexpectedly, the
shapes of the luminosity functions for the resolved stars in M31 and
M32 are similar (despite significant differences in optical absorption
line spectra).  M32 differs only in having fewer total stars (per unit
UV surface brightness) above the detection threshold.  Brown and
coworkers also find that about 10\% of the brighter resolved
population is not explainable by existing post-HB evolutionary
tracks.

Any process that reduces HB envelope masses in a significant fraction
of the population can have influence extending beyond the UV region.
EHB stars do not become AGB stars, and if most of the evolving
population passes through the small-envelope channel, the AGB
contribution to the integrated optical/IR spectrum (mainly longward of
6000 \AA) of a galaxy will decrease.  Changes in the AGB should also be
detectable with the surface brightness fluctuation imaging method
(Tonry \& Schneider 1988).  Ferguson \& Davidsen (1993) find that the
incidence of planetary nebulae, which should also decrease if the AGB
population decreases, is anticorrelated with bluer 1500--V colors.
This is important circumstantial evidence that small-envelope HB stars
are implicated in the UVX.  The correlation should be pursued with a
larger sample of galaxies, and radial dependences within galaxies
should be studied as well.  

The existing deep imaging studies therefore provide good support for
the EHB interpretation of the UVX.  Imaging to the level of the EHB
itself within the Local Group can probably be secured with HST/STIS
and HST/ACS.  Worthey (1993) has described how the surface brightness
fluctuation technique can be applied to faint hot stars to extend the
effective depth of such UV imaging.

\subsection{\it Cosmic Evolution of the UVX}

Since the models predict that the UVX (if dominated by the EHB or PAGB
channels) should decrease as the main-sequence turnoff mass increases,
there should be strong evolution of the UVX with lookback time (see \S 7.2).  If E
galaxies are sufficiently homogeneous, there could be a unique
lookback beyond which the UVX disappears.  Given the uncertainty in
the models discussed above, lookback effects should probably be viewed
for the present more as a valuable opportunity to refine the models
than as a way to age-date the universe.  

There are serious technical challenges in making restframe UV
measurements at moderate redshifts.  The galaxies are faint.  At a
redshift of 0.5 (a lookback time of 6 Gyr), the distance modulus is
43, implying that the unevolving UVX of a strong upturn source in a
typical luminous elliptical would have $m_{\lambda}(2250\, {\rm \AA})
\sim 24.5$.  Simple detection of far-UV light (e.g.\ in broad bands)
is not sufficient to distinguish a UVX component from the decaying
initial burst or late star formation (see \S 5).  Multiband photometry
or spectroscopy is necessary.  Several attempts to observe the UVX at
high redshift have been made (e.g.\ Windhorst et al 1994), but only
recently has a detection been claimed in the cluster Abell 370 ($z =
0.38$) by Brown et al (1998b).  Using broad-band filters with HST/FOC,
they find four cluster E galaxies to have a range of 1500--V similar
to that in local galaxies.  If this is UVX light, it implies a high
formation redshift ($z_{\rm F} > 4$) in the context of most existing
models.  An absence of UV evolution over the past few Gyr would be
inconsistent with some classes of UVX models.  It will be especially
important to link changes in the UVX of distant galaxies with
evolution of the initial burst at optical/IR wavelengths (now detected
up to $z \sim 0.9$, e.g.\ Stanford et al 1998).

\subsection{\it Summary and Key Issues}

Progress on the UVX problem during the last ten years has been
excellent.  The theoretical, spectral, and imaging evidence has
recently converged toward the view that the UVX originates from
He-burning, extreme horizontal branch stars, their post-HB progeny,
and post-AGB stars in the dominant, metal-rich stellar population of E
galaxies.  The mixture of these types apparently varies from object to
object, perhaps in a systematic way with global mean metallicity or
mass, but in most cases the EHB/post-EHB channels are the more
important.  The simplest explanation for the correlation between the
UVX and optical line strengths is that the mass-loss parameter
\etar\ increases with $Z$ or that $\Delta Y/\Delta Z \ga 2.5$.

Although evolutionary synthesis models successfully predict UV
spectral properties in the ranges observed, progress in understanding
the UVX, and in refining estimates of ages and abundances derived
therefrom, is hampered by our lack of knowledge of two basic
processes:  mass loss on the giant branch and helium enrichment.  Both
of these are critical to the efficiency with which an old population
can generate UV-bright stars.  We urgently require a more complete and
predictive physical theory of giant-branch mass loss.  This is the
highest priority for UVX theory in the near term.  The question of the
value of the helium enrichment parameter ($\Delta Y/\Delta Z$) near
and above solar abundance also needs to be addressed.  Both areas
demand extensive observational programs on nearby systems as well as
fundamental improvements in theoretical modeling.  The same is true of
diffusion in hot atmospheres, which is important to interpreting the
UVX line spectrum.

These are the most serious gaps in our astrophysical understanding of
the UVX, but there are other troublesome issues as well, three of
which are worth mentioning:

\noindent 1. The behavior of the UVX seems to be firmly linked to that
of the lighter elements such as N, Mg, and Na and decoupled from the
Fe-peak (\S 5.2).  This adds an additional dimension to modeling
space, so far unexplored, which is not at present well supported by
nucleosynthetic theory (Worthey 1998).  

\noindent 2. The internal spatial gradients in 1500--B color
discussed in \S 4.2 do not correlate with gradients in \mgii\ (Ohl et
al 1998).  Metallicity is evidently not the sole parameter governing
the UVX.  This may be related to the decoupling of the Fe-peak noted
in paragraph 1 above, or it may reflect the influence of other changing
parameters within galaxies, such as age or $Y$ abundance.  M32, with a
large and reversed UVX gradient (see Figure 3), is an important case
since there is considerable independent evidence for an intermediate
age ($\la 8$ Gyr) population there and possibly an age gradient in
which the central regions are younger (O'Connell 1980, Freedman 1992,
Rose 1994, Hardy et al 1994, Faber et al 1995, Grillmair et al 1996).

\noindent 3. There has been very little work on the dependence of the
UVX on galaxy morphology despite suggestions of differences between E
galaxies and S0 galaxies (e.g.\ Smith \& Cornett 1982).  Bright,
nearby spiral bulges could readily be studied in the UV with HST, and
comparisons with E galaxies could help distinguish some of the
underlying drivers of the UVX phenomenon.  If, for example, bulges
have a wider range of ages than E galaxies (e.g.\ Wyse et al 1997),
then the younger ones should have smaller UV upturns than E's.

\section{OTHER FAR-UV PHENOMENA IN E GALAXIES}

Although the UVX produces a ubiquitous extended light background in old
populations, it is at a low level and is coincident with a ``dark
window'' in the natural sky background centered at about 2000 \AA\,
where the sky is about 40$\times$ fainter than at any other wavelength
in the optical-IR region (O'Connell 1987).  The faint UV backgrounds
permit isolation of other interesting phenomena that are either
unique to the UV or are drowned out in the visible bands by the glare
of the main sequence and giant branch stars.  This includes
low-luminosity active nuclei, recent star formation, blue straggler
populations, gas in the $10^5$--$10^6$ K temperature range, scattered
light from dust grains, and H$_2$ fluorescence features near 1600 \AA.
As noted in \S 3, hot continuum sources that contribute as little as
0.1\% of the V-band light of a galaxy can be readily detected in the
UV region.   In this section we briefly discuss UV
observations relevant to massive star formation and nonthermal nuclei.

\noindent{\it Recent Massive Star Formation.} Identification of a
minority component of massive stars in an old population depends on
detection of spectral or color distortions in integrated light or on
imaging of individual stars or concentrations of stars that stand out
against the smooth background light.  The vacuum UV is about
30--50$\times$ more sensitive to such effects than the optical/IR bands
(McNamara \& O'Connell 1989).  Based on spectral synthesis models for
constant star formation with a normal IMF (e.g.\ Bruzual \& Charlot
1993, Cornett et al 1994), the star formation rate per unit V-band
luminosity is related to far-UV color as follows:  $$
\dot{S}/L_{V} \sim 8 \times 10^{-11}\, 10^{-0.4(1500-V)} {\msunyr\,
L_{V,\odot}^{-1} }.$$  The 1500 \AA\ flux used to compute the color
here is the part of the total far-UV flux that is attributed to
young stars.  The coefficient in this expression is almost independent
of the period over which the star formation is assumed to have
persisted, for periods over 50 Myr.  

If all of the UV light in the strongest UV upturn cases (1500--V
$\sim$ 2) were attributed to massive star formation, the implied
normalized rate would be $ \dot{S}/L_{V} \sim 1.3 \times
10^{-11}$, or a total rate of $\dot{S} \sim 0.25 \msunyr$ for a
typical gE galaxy with $M_V = -21$.  This is obviously a strong upper
limit to massive star formation in a normal E galaxy since only a
small fraction of the far-UV light can be produced by massive stars,
as discussed in detail in \S 4 and \S 5.  If we take 20\% as the upper
limit on the contribution of young starlight at 1500 \AA\ in a galaxy
with a more typical UV upturn with 1500--V = 3.5, then the maximal
total rate becomes $\dot{S} \la 0.01 \msunyr$.  This is a very
stringent limit on the amount of continuing star formation in a
typical gE galaxy.

These values can be compared with the estimated total mass loss from
stars evolving up the giant branch.  The ``evolutionary rate'' in an
old population (i.e.\ the number of stars evolving off the main
sequence per unit time) is $\sim 4 \times 10^{-11}\, {\rm yr}^{-1}\,
L_{V,\odot}^{-1}$ (e.g.\ Renzini \& Buzzoni 1986, DOR).  If each star
sheds 0.3--0.5 \msun, then the total estimated normalized mass loss
rate is $\dot{m}/L_{V} \sim 1$-2$ \times 10^{-11} {\msunyr\,
L_{V,\odot}^{-1} }$, or 0.2--0.4 \msunyr\ for a galaxy with $M_V = -21$.
The maximal continuing star formation rate derived from far-UV
data is some 20--40$\times$ smaller.  Clearly, most of the material
produced by giant branch mass loss is not being recycled into new
stars in normal E galaxies, at least not with a normal IMF.  The UV is
the key to this conclusion, since high S/N optical-band studies
generally cannot exclude complete recycling (e.g.\ O'Connell 1980,
Gunn et al 1981).  

The ultimate fate of the lost red giant envelopes remains unclear.  At
early times the material is probably removed from galaxy interiors by
high-temperature, supernovae-driven winds.  In more massive galaxies,
the gas forms a hot corona, which is detectable at X-ray wavelengths
(e.g.\ Forman et al 1985).  Some fraction of the corona is returned to
the interior by a cooling flow (e.g.\ Sarazin \& White 1988, David et
al 1991), but the final repository of the material from the flow remains
to be identified.  One interesting example of young stars
in a normal old population is the remarkable source P2, which
is coincident with the dynamical center of M31 (King et al 1995, Lauer
et al 1998).  This is slightly extended and considerably  bluer than the
surrounding UVX population.  It has the characteristics of an
intermediate-age star cluster, but with $M_V = -5.7$, it can account for
only a tiny fraction of recent mass loss by the bulge giants.  Its
massive stars may have been formed through stellar collisions.  
The second concentrated nuclear source in M31, denoted P1, is not
at the dynamical center and is brighter at optical wavelengths.  However,
its UV properties are similar to those of the inner bulge.  It has
been suggested that this is a cannibalized galaxy nucleus in the
final stages of consumption by M31.  If so, it has managed to clothe
itself with a UVX population similar to the bulge stars in M31.  

The minority of nearby early-type galaxies that do exhibit evidence
for recent star formation (including NGC 205, 5102, and 5253) have
probably mostly suffered gas transfer during a recent interaction.  UV
observations in these cases provide a much improved picture of the
massive star population and its history than do optical data (e.g.\
BBBFL, Wilcots et al 1990, Deharveng et al 1997, Calzetti et al 1997).
By contrast with normal E galaxies, recent star formation is
often found in early-type galaxies associated with massive cluster
cooling flows (reviewed in Fabian 1994).  Systems with UV observations
include M87, Abell 2199, and NGC 1275 (Perola \& Tarenghi 1980,
Bertola et al 1982, Bertola et al 1986, BBBFL, McNamara \& O'Connell
1989, Smith et al 1992, Dixon et al 1996).  The UV is important here
in placing better limits on star formation rates (always much smaller
than X-ray estimates of total accretion rates) and in exploring
possible anomalies in the initial mass function.

\noindent{\it Active Nuclei.} The flat energy distributions of nuclear
nonthermal sources imply that the contrast of an AGN against its
surroundings in an E galaxy can improve by a factor up to $\sim 100$
in the UV compared with the optical-IR.  This permits better study of
known nuclei and searches for very low-luminosity activity.  A number
of identifications of nuclear point sources have recently been made by
UV imaging either of complete samples of nearby galaxies (Maoz et al
1995, 1996) or of samples of objects with Low Ionization Nuclear
Emission Region (LINER) optical spectra (Barth et al 1998).  Only
about 30\% of the known LINERs are detected this way, and Maoz and
Barth and their respective colleagues suggest that obscuration by dust
reduces the visibilty of the other nuclei, at least in the disk
galaxies in their samples.  However, the UV brightnesses of the
nonthermal nuclei support photoionization (rather than shock
excitation) models for the LINER emission lines.  

In the case of E galaxies with known bright nuclei (e.g.\ M87), the
AGN contributes only a small part of the FUV light within the IUE
aperture.  From the UIT images of 
Ohl et al (1998), we find that the
nucleus and jet in M87 produce only 10\% of the FUV light within a
radius of 10\arcsec.  A similar situation applies to NGC 4278, whose
nonthermal nucleus was recently detected by Moller et al (1995).
These amounts are, however, sufficient to shift the active galaxies
such as M87, NGC 4278, and NGC 1052 slightly in 1500--V vs \mgii\
diagrams such as Figure 6 (as first remarked by BBBFL).

The most interesting case of UV-facilitated observations of an E
galaxy AGN is that of NGC 4552.  This object has conspicuous radio and
infrared signatures of an active nucleus and was originally observed
with IUE for that reason (O'Connell et al 1986).  Aside from a strong,
spatially-extended UV-upturn, however, there were no nuclear anomalies
obvious until HST imaging was obtained by Renzini et al (1995) and
Cappellari et al (1998).  The HST observations show a time-variable,
unresolved ($r \la 0.07\arcsec$) spike of UV light which brightened by
a factor of 4.5$\times$ between 1991 and 1993.  Without the resolution
of HST and the improved contrast offered by the UV, it would have been
impossible to detect this source, which is currently the least
luminous known AGN, having an H$\alpha$ luminosity of only $6 \times
10^{37} {\rm erg}\, {\rm s}^{-1}$.  The outburst probably corresponds
to the accretion of material stripped from a single star during a close
fly-by of the nuclear black hole (Cappellari et al 1999).

\section{CONCLUSION}

Ultraviolet observations have opened a new, and unexpectedly rich,
window on old stellar populations that has revealed phenomena that
are either difficult or impossible to study at longer wavelengths.
The identification of the UVX component with low-mass, small-envelope
stars has led to the recognition that the spectra of distant E
galaxies are remarkably sensitive to what in traditional stellar
population research would have been regarded as subtle astrophysical
processes, including giant-branch mass loss, helium enrichment, and
atmospheric diffusion.  The fact that these processes are manifestly
not properly understood at the moment, precluding a definitive
interpretation of the UVX in terms of global population parameters,
is less important than the long-term promise of UV observations as
powerful probes of galaxy evolution.

\acknowledgments

\noindent {\it Acknowledgements:} For comments, figures, and other
help in preparing this paper, I am most grateful to Ralph Bohlin, Tom
Brown, Dave Burstein, Daniela Calzetti, Jeff Crane, Ben Dorman, Harry
Ferguson, Ian Freedman, Richard de Grijs, Wayne Landsman, Ray Ohl,
Alvio Renzini, Bob Rood, Ted Stecher, and Sukyoung Yi.  This work has
been supported in part by NASA Long Term Space Astrophysics grant
NAG5-6403.


\def \aj             {{\it Astron.~J.}}
\def \apj            {{\it Ap.~J.}}
\def \apjlett        {{\it Ap.~J. Lett.}}
\def \apjsupp        {{\it Ap.~J. Suppl.}}
\def \apspsci        {{\it Astrophys.~Space Sci.}}
\def \astrap         {{\it Astron.~Astrophys.}}
\def \astraplett     {{\it Astron.~Astrophys.~Lett.}}
\def \astrapsupp     {{\it Astron.~Astrophys.~Suppl.}}
\def \mnras          {{\it MNRAS}}
\def \pasp           {{\it Publ.~Astron.~Soc.~Pac.}}
\def \sci            {{\it Science}}
\def \nature         {{\it Nature}}
\def \annrev         {{\it Annu. Rev. Astr. Ap.}}

\end{document}